\documentclass{jfm}

\usepackage{epsfig}
\usepackage{graphicx}
\usepackage{natbib}
\usepackage{pdflscape}
\usepackage{bm}
\usepackage{float}
\usepackage{color}

\def\Rk{\mbox{\rm R}_\kappa}

\def\Rey{\mbox{\rm Re}}
\def\cs{c_{\rm s}}

\newcommand{\FF}{\mbox{\boldmath $F$} {}}
\newcommand{\ff}{\mbox{\boldmath $f$} {}}

\newcommand{\be}{\begin{equation}}
\newcommand{\ee}{\end{equation}}
\newcommand{\bea}{\begin{eqnarray}}
\newcommand{\eea}{\end{eqnarray}}

\newcommand{\nab}{\mbox{\boldmath $\nabla$} {}}
\newcommand{\DD}{{\rm D} {}}

\newcommand{\Eq}[1]{Eq.~(\ref{#1})}
\newcommand{\Eqs}[2]{Eqs~(\ref{#1}) and~(\ref{#2})}

\newcommand{\bra}[1]{\langle #1\rangle}
\newcommand{\bbra}[1]{\left\langle #1\right\rangle}

\newcommand{\kk}{\mbox{\boldmath $k$}}
\newcommand{\pp}{\mbox{\boldmath $p$}}

\newcommand{\aaa}{\mbox{\boldmath $a$}}
\newcommand{\AAA}{\mbox{\boldmath $A$}}
\newcommand{\q}{\mbox{\boldmath $q$}}
\newcommand{\x}{\mbox{\boldmath $x$}}
\newcommand{\y}{\mbox{\boldmath $y$}}
\newcommand{\rr}{\mbox{\boldmath $r$}}

\newcommand{\rvu}[1]{\hat{r}_{#1}}

\newcommand{\PPP}[1]{\tilde{P}_{#1}}
\newcommand{\uu}{\mbox{\boldmath $u$}}
\newcommand{\RR}{\mbox{\boldmath $R$}}

\newcommand{\xo}{\mbox{$\x_0$}}

\newcommand{\yo}{\mbox{$\y_0$}}

\newcommand{\ro}{\mbox{$\rr_0$}}

\shorttitle{Passive scalar mixing and decay in renewing flows} 
\shortauthor{Aiyer, Subramanian and Bhat} 

\title{Passive scalar mixing and decay at finite correlation times 
in the Batchelor regime}

\author
 {
 Aditya K Aiyer\aff{1,2,3}
  \corresp{\email{aditya.kandaswamy@gmail.com}},
  Kandaswamy Subramanian\aff{4}
  \and 
  Pallavi Bhat\aff{4,5,6}
  }

\affiliation
{
\aff{1}
BITS-Pilani, K. K. Birla Goa campus, Zuarinagar, Goa-403726, India
\aff{2}
TIFR Centre for Interdisciplinary Sciences
Narsingi, Hyderabad 500075, India
\aff{3}
Department of Mechanical Engineering,
The Johns Hopkins University, Baltimore, MD 21218, USA
\aff{4}
IUCAA, Post Bag 4, Ganeshkhind, Pune 411007, India
\aff{5}
Department of Astrophysical Sciences and
Princeton Plasma Physics Laboratory,
Princeton University,
Princeton, NJ 08543, USA
\aff{6}
Plasma Science and Fusion Center, Massachussetts Institute of Technology, Cambridge, MA 02139, USA.
}

\begin{document}

\maketitle

\begin{abstract}

An elegant model for passive scalar mixing  and decay was given 
by \citet{K68} assuming the velocity to be delta-correlated in time.
For realistic random flows this assumption becomes invalid.
We generalize the Kraichnan model to include the effects of a finite correlation time, $\tau$, using renewing flows. 
The generalized evolution equation for the 3-D passive scalar spectrum $\hat{M}(k,t)$ or its correlation function $M(r,t)$,  
gives the Kraichnan equation when $\tau \to 0$, and extends it to
the next order in $\tau$. It involves third and fourth order derivatives 
of $M$ or $\hat{M}$ (in the high $k$ limit). 
For small-$\tau$ (or small 
Kubo number), it  can be recast using the Landau-Lifshitz approach, to one
with at most second derivatives of $\hat{M}$.
We present both a scaling solution to this equation 
neglecting diffusion and a more exact solution including
diffusive effects.  
To leading order in $\tau$, 
we first show that the steady state 1-D passive scalar spectrum,
preserves the \citet{B59} form, $E_\theta(k) \propto k^{-1}$,
in the viscous-convective limit, independent of $\tau$.
This result can also be obtained in a general manner using
Lagrangian methods. Interestingly, in the absence of sources,
when passive scalar fluctuations decay, we show that 
the spectrum in the Batchelor regime and at late times,
is of the form
$E_\theta(k) \propto k^{1/2}$  and also  independent of $\tau$.
More generally, finite $\tau$ does not qualitatively change the
shape of the spectrum during decay.
The decay rate is however reduced for finite $\tau$. 
We also present results from high resolution ($1024^3$) direct numerical simulations
of passive scalar mixing and decay.
We find reasonable agreement with predictions of the Batchelor
spectrum, during steady state. The scalar spectrum during
decay is however dependent on initial conditions.
It agrees qualitatively with analytic predictions when power is dominantly in wavenumbers corresponding to the
Batchelor regime, but is shallower when box scale fluctuations dominate during decay.
\end{abstract}

\section{Introduction}
Understanding turbulent mixing and decay of passive scalars is important in a number of natural settings, for many practical applications and 
also as a means to get insight into turbulence itself \citep{SS00,W00,SS10,DKS13}. These settings and applications 
include atmospheric phenomena,  oceanography, combustion, dispersion of pollutants, transport
of heat and even metal mixing in the interstellar medium of galaxies.
Progress in understanding passive scalar turbulence may also 
ultimately lead to a better understanding of fluid turbulence itself,
as it offers a simpler setting wherein many of the issues of fluid turbulence can be examined.

The spectrum of passive scalar fluctuations advected by
a turbulent or random flow has received attention right from the early days
of turbulence research \citep{O49,C51,B59,K68}.
This spectrum depends on the relative importance
of the viscous dissipation (as characterised by the kinematic viscosity $\nu$)  
compared to the scalar diffusivity ($\kappa$).
If the fluid Reynolds number, $\Rey=u/(k_f\nu)$ is large such that there
is an extensive inertial range for the turbulent velocity, and
the corresponding Peclet number $\Rk=u/(k_f\kappa)$ is also comparable,
the one dimensional scalar spectrum is expected to scale as,  
$E_\theta(k) \propto k^{-5/3}$, the same form as the Kolmogorov spectrum
for the velocity \citep{O49,C51,B59}. 
Here $u$ and $k_f$ are characteristic velocity and
wavenumber where the fluid is driven in some random fashion. 
This range of wavenumbers is referred to as the inertial-convective
range, where the velocity fluctuations belong to the inertial range
and convect the passive scalar.

Another interesting regime, often referred to as the viscous-convective or the
Batchelor regime, obtains when the Schmidt number $Sc = \nu/\kappa \gg1$
(or the thermal Prandtl number when the passive scalar is temperature).
In this case the viscous cut-off scale is much larger than that due to
the scalar diffusion; or the corresponding viscous cut-off wavenumber 
$k_\nu \ll k_\kappa$,
the cut-off wavenumber corresponding to the scalar fluctuations. 
The velocity field below the viscous scale is smooth and
thus the scalar mixing in this regime is dominated by the random shearing 
by relatively smooth viscous scale eddies. 

The resulting passive scalar
spectrum was analysed by \citet{B59} (B59), 
based on the effect of linear shear flows on the scalar density.
For an incompressible velocity ${\bf u}$, a given scalar blob 
is eventually stretched along one direction and compressed along the
direction of the maximally negative strain. 
B59 assumed that the principal axes of the flow are 
random and isotropically distributed, but the magnitudes of the strains
are fixed. On averaging over the directions of
these principal axes, B59 then argued that
the steady state 1-D scalar spectrum $E_\theta(k) \propto 1/k$, 
for $k_\nu \ll k \ll k_\kappa$. In the above analysis, the randomness
in time of the velocity field, that it has a finite correlation time
in a realistic turbulent flow, was not explicitly taken into account.

\citet{K68} (K68) re-examined the problem of passive scalar mixing 
and decay in the opposite extreme case of a delta-correlated in time velocity field. 
This assumption of delta-correlation, allows
one to convert the stochastic differential equation for the passive scalar,
to a partial differential equation in real space for the time
evolution of the scalar correlation function $M(r,t)$.
In Fourier space, such an equation becomes an integro-differential
equation for the 3-D scalar spectrum $\hat{M}(k,t)$. 
However, in the viscous-convective 
limit when, $k_\nu \ll k \ll k_\kappa$, one can make a small
$r$ expansion, and the Fourier space
evolution equation also becomes a differential equation. 
\citet{K68} deduced that, a steady rate of transfer of 
the scalar variance from 
wavenumbers below $k$ to those above, requires 
$E_\theta(k) =4\pi k^2 \hat{M}(k) \propto k^{-1}$, or the Batchelor spectrum.
Such a steady state can be 
obtained when there is a source of scalar fluctuations
at small $k$. The Kraichnan assumption also allows one to
deduce many other interesting results on higher order correlations 
\citep{FGV01}.
The Batchelor spectrum was also argued for, in general terms by 
\citet{G68}.
A useful review of the Batchelor spectrum in isotropic turbulence is given
in \citet{DSY10} (see also the chapter by \citet{GY13}).

There has also been extensive work on general aspects of 
scalar mixing in the Lagrangian
framework \citep{K74,CFKL95,CFL96,BF99,FGV01,CEFV01,FL03,V06,CKL07}. 
In the presence of a source for the scalar, 
one can relate the two-point spatial
correlator of the passive scalar, to the probability $P$, of
the two points having a separation $r_0$ at an earlier time $t_0$,
given that they have a separation $r$ at time $t$.
Once simplifying assumptions about the source correlator and the
probability $P$ are made, 
the Batchelor spectrum or the corresponding 
two-point spatial correlator can be derived, in a manner which does not 
explicitly involve the correlation time $\tau$
(see \citet{FGV01} for a review and Appendix~\ref{appA}).

Another important aspect of passive scalar evolution in random flows
is what happens in the absence of a source. The passive scalar 
fluctuations are expected to decay in such a case. The scalar
spectrum and its properties during this decay phase 
are also of interest, especially 
when the velocity field has a finite correlation time.
This aspect is more difficult to handle, and
does not appear to have received as much attention as
the steady state case.
In the Lagrangian approach, it requires an explicit calculation of $P$
which is 
non-trivial.
Studying finite $\tau$ effects on the passive scalar spectrum during decay,
and the decay rate,
also forms one of the motivations of the current work. 

The analysis of \citet{B59} 
or the Lagrangian framework, by their very nature, do not yield 
a differential equation for the time
evolution of the scalar spectrum
(or correlation function), for any finite $\tau$.
At the same time the assumption by \citet{K68}, of delta-correlated in
time velocity field is not realistic in turbulent fluids. There the correlation
time, $\tau$, is expected to be at least of the order of 
the smallest eddy turn over time.
Thus, it is important to systematically extend the K68 analysis to take 
into account the effects of a finite correlation time, and 
derive a corresponding evolution equation which extends the 
Kraichnan equation to finite $\tau$. It would also be of interest
to then examine quantitatively, how this alters the evolution of
the scalar fluctuations.
Firstly, if indeed  the Batchelor spectrum arises as a steady state solution of this extended Kraichnan equation. 
Also importantly, in the case of decay due to absence of sources, 
how the scalar spectrum and decay rate
change due to finite $\tau$ effects?
These issues form the main motivation of the present work.

For this purpose, we use what are known as renewing (or renovating) flows, 
which are random flows where the velocity field is renewed after 
every time interval $\tau$. Such flows have been used by 
\citet{Zeld88b}, to discuss both the diffusion of scalars, 
in limit of $\tau \to 0$, and the corresponding magnetic field
generation problem; first worked out by \citet{Kaz68} also
in the delta-correlated in time limit. 
They have also been used to study the effect of finite
correlation time on mean field dynamos \citep{DMSR84,GB92,KSS12},
and fluctuation dynamos \citep{BS14,BS15}. The work by
\citet{BS14,BS15} in particular, extended the \citet{Kaz68} evolution
equation for the magnetic field correlator,
to finite correlation times. It also showed that,
what is known as the Kazantsev spectrum for the magnetic field,
(which obtains for delta-correlated in time flows), is
preserved even at a small but finite $\tau$.
We will use here some of the methods developed by \citet{BS15} (BS15) for the fluctuation dynamo
problem and apply them to passive scalar mixing
and decay.
 
In the next section, we formulate the analysis of passive scalar mixing 
and decay
in renewing flows. The detailed derivation of the Kraichnan 
evolution equation for $M(r,t)$ and $E_\theta(k,t)$, 
and also its generalization to incorporate finite-$\tau$ effects 
to the leading order, is given in section 3. Analysis of this 
generalized Kraichnan evolution equation for both the steady state, and in terms of a Green's function
based solution for the decay, is done in section 4.
A brief discussion of passive scalar mixing using  
Lagrangian methods is given in Appendix~\ref{appA}.
Results from direct numerical simulations to test the analytical predictions
are presented in section 5. We end with a discussion
of our results in the last section.
\section{Passive scalar evolution in renewing flows}
\label{sect2}
The evolution of a passive scalar field $\theta(\x,t)$, advected
by a fluid with velocity $\uu$, is given by,
\begin{equation}
\frac{\partial \theta}{\partial t} + \uu\cdot{\bf \nabla}\theta =
\frac{D\theta}{Dt} = \kappa\nabla^2 \theta,
\label{scalareqn}
\end{equation}
where we assume the motions to be incompressible
with ${\bf \nabla}\cdot\uu =0$. Also $D\theta/Dt$ is the
Lagrangian time derivative of the passive scalar along the
fluid trajectory. We adopt here a velocity field that is random and renews
itself every time interval $\tau$ \citep{DMSR84,GB92}. We take
$\uu$ to be the form given 
by \citet{GB92}(GB), 
\begin{equation}
\uu(\x)=\aaa\sin(\q\cdot\x+\psi).
\label{uturbdef}
\end{equation}
The incompressibility condition puts an additional
constraint $\aaa\cdot\q=0$.
In each time interval $\left[(n-1)\tau, n\tau\right]$,\\
(i) $\psi$ is chosen uniformly random between 0 to $2\pi$, \\ 
(ii) $\q$ is uniformly distributed 
on a sphere of radius $q=\vert\q\vert$, and \\
(iii) for every fixed $\hat{\q}=\q/q$, the direction of 
$\aaa$ is uniformly distributed 
in the plane perpendicular to $\q$.\\
Condition (i) on $\psi$ ensures statistical homogeneity,
while (ii) and (iii) ensure statistical isotropy
of the flow. Further, for ease of computations, 
we modify the GB ensemble and use \citep{BS14,BS15},
\be
a_i = \PPP{ij} A_j, ~~\PPP{ij}(\hat{\q}) = \delta_{ij} - \hat{q}_i\hat{q}_j
\label{aEns}
\ee
where $\AAA$ is uniformly distributed on a sphere of radius $A$, and
$\PPP{ij}$ projects $\AAA$ to the plane perpendicular to $\q$.
Then on averaging over $a_i$ and using the fact that $\AAA$
is independent of $\q$, we have $\bbra{\uu}=0$ and,
\be
\bra{a_ia_l}=\bra{a^2}\frac{\delta_{il}}{3}=\bbra{A_jA_k\PPP{ij}\PPP{lk}}=A^2\frac{\delta_{jk}}{3}\bbra{\PPP{ij}\PPP{lk}} 
=\frac{A^2}{3}\bbra{\PPP{il}}=\frac{2A^2}{3}\frac{\delta_{il}}{3}
\label{aArel}
\ee
and so $\bra{a^2} = 2A^2/3$.
This modification in ensemble
does not affect any result using the renewing flows.

Note that we could also add a random forcing term, $f_\theta(\x,t)$, to 
\Eq{scalareqn} governing the evolution of the passive scalar, so as
to provide a source to scalar fluctuations, at low 
wavenumbers \citep{SS00,FGV01}. 
In general $f_\theta$ is assumed to be statistically stationary, homogeneous, 
isotropic, and Gaussian random with zero mean. It is also often idealized to
have short (or even delta) correlation in time, and with a prescribed 
Fourier space spectrum $F(k/k_L)$, which is peaked around a 
sufficiently low wavenumber $k_L$, for example the wavenumber where
velocity fluctuations are also injected.
One useful case is when $f_\theta$ is also renewed randomly every
time interval $\tau$ independent of $\theta$ during previous
times. In Fourier space, such a term provides a source of fluctuations 
at small wavenumbers \citep{K68,SS00}
outside the range of interest in this paper, 
but which is important if one wants to maintain a steady state. 
This is discussed further in Appendix~\ref{appA}
when reviewing a Lagrangian analysis of the steady state Batchelor spectrum.
In the rest of the main text 
however,
we do not explicitly consider the effect of $f_\theta$.

In any time interval $\left[(n-1)\tau, n\tau\right]$, the 
passive scalar evolution is given by
\be
\theta(\x,n\tau) \;=\; \int G(\x,\xo,\tau)
\theta(\xo,(n-1)\tau) \ d^3\xo
\ee
where $G(\x,\xo)$ is the Green's function of 
Eq.~(\ref{scalareqn}).
The passive scalar two-point spatial correlation function is defined as
\be
\bra{\theta(\x,t)\theta(\y,t)} = M(r,t), \quad {\rm where}
\quad r=\vert\rr\vert = \vert(\x -\y)\vert,
\label{scalarcor}
\ee
and $\bbra{.}$ denotes an ensemble average.
Here the passive scalar is assumed to be statistically homogeneous 
and isotropic. Note that this feature is preserved by renewing flow
if initially the scalar field is statistically 
homogeneous and isotropic (see below).
Then the evolution of the scalar fluctuations is governed by
\be
M((\x-\y),n\tau)
= \int \tilde{\mathcal{G}}(\x,\xo,\y,\yo,\tau)
M((\xo-\yo),(n-1)\tau) \ d^3\xo \ d^3\yo.
\label{greenScal}
\ee
Here $\tilde{\mathcal{G}}=\bbra{G(\x,\xo,\tau)G(\y,\yo,\tau)}$. 
It is important to note that, 
we could split the averaging on the right side of equation
between the Green's function and the initial scalar correlator,
because the renewing nature of flow implies that the Green's function in the
current interval is uncorrelated to the scalar
correlator from the previous interval. 
The renewing nature of the flow also implies that $\tilde{\mathcal{G}}$ 
depends only on the time difference, $\tau$ and not separately 
on the initial and final times in the interval $[(n-1)\tau,n\tau]$. 

We use the method of operator splitting introduced by GB in order to calculate 
$\tilde{\mathcal{G}}(\x,\xo,\y,\yo,\tau)$ in the renewing flow (see also
\citet{holden}).
A more detailed discussion of this method, can also be found in 
\citet{BS15} (BS15).
The renewal time, $\tau$, is split into two 
equal sub-intervals. In the first sub-interval $\tau/2$, we consider 
the scalar to be purely advected with twice the original velocity, 
neglecting the diffusive term, that is setting $\kappa=0$ in \Eq{scalareqn}. 
In the second sub-interval, $\uu$ is 
neglected and the field is diffused with twice the value of the original diffusivity. 

This operator splitting method, plausible in the small $\tau$ limit,
has been used earlier to recover both the standard mean field dynamo
and fluctuation dynamo evolution equations in renewing
flows \citep{GB92,KSS12,BS14,BS15}. It is also further discussed in 
the BS15, where it is shown that the above two
operations commute for the magnetic correlator in the small diffusivity 
limit, which we shall assume here as well.

In the first sub-interval $\tau/2=t_1-t_0$, 
taking $\kappa=0$, the advective part of Eq.~(\ref{scalareqn}), 
is simply $D\theta/Dt =0$. Thus $\theta$ is constant along
the Lagrangian trajectory of the fluid element or
\be
\theta(\x,t_1) = \theta(\xo,t_0).
\label{cauchy}
\ee
Here $\theta(\xo,t_0)$ is the initial field, 
which is simply advected 
from $\xo$ at time $t_0$, to $\x$ at time $t_1 = t_0 +\tau/2$.  

Note that the phase $\Phi = \q\cdot\x +\psi$
in Eq.~(\ref{uturbdef}), 
is constant in time as $d\Phi/dt = \q\cdot\uu =0$, 
from the condition of incompressibility. 
Thus $d\x/dt = 2\uu$ can be integrated to obtain,
at time $t_1 = t_0 +\tau/2$, 
\be
\x = \xo +\tau \uu = \xo + \tau {\bf a} \sin(\q\cdot\xo + \psi).
\label{traj}
\ee
It will be more convenient to work with 
the resulting scalar field in Fourier space, 
\be
\hat{\theta}({\bf k},t_1) = \int \theta(\x,t_1) e^{-i \kk\cdot\x } d^3 \x
= \int \theta(\xo,t_0) e^{-i \kk\cdot\x } d^3 \x.
\label{adveceq}
\ee
Next, in the second sub-interval ($t_1,t=t_1+\tau/2$), 
only diffusion operates with diffusivity $2\kappa$ to give,
\be
\hat{\theta}(\kk,t) = G^{\kappa}(\kk,\tau)\hat{\theta}( \kk,t_1) 
= e^{-(\kappa \tau \kk^2)} \hat{\theta}( \kk,t_1).
\label{diffeq}
\ee
Here $G^{\kappa}$ is the diffusive Green's function.

\subsection{Evolution of passive scalar fluctuations}

Let us assume that the mean scalar field vanishes.
In order to derive the evolution equation for the 
scalar two point correlation function,
we combine \Eq{adveceq} and \Eq{diffeq} to get,
\be
\bra{\hat{\theta}(\kk, t)\hat{\theta}^*(\pp, t)} 
= e^{-\kappa\tau(\kk^2+\pp^2)}\int 
\bra{e^{-i(\kk\cdot\x-\pp\cdot\y)}} M(\ro,t_0) d^3\x d^3\y,
\label{corrlmaineq1}
\ee
where we have defined $\ro=\xo-\yo$.
The statistical homogeneity of the field, allows us to write
the two-point scalar correlator in Fourier space as, 
\be
\bra{\hat{\theta}(\kk, t)\hat{\theta}^*(\pp, t)} 
=(2\pi)^3 \delta^3(\kk-\pp) 
\hat{M}(\kk,t).
\label{fouriercor}
\ee
We now follow a method similar to that used in BS15.
We can transform $(\x,\y)$ in Eq.~(\ref{corrlmaineq1}) to $(\xo,\yo)$
using Eq.~(\ref{traj}). 
The Jacobian of this transformation 
is unity due to incompressibility of the flow.
Then we write
$\kk\cdot\xo - \pp\cdot\yo = \kk\cdot\ro + \yo\cdot(\kk-\pp)$ 
in Eq.~(\ref{corrlmaineq1}), transform from $(\xo,\yo)$ 
to a new set of variables $(\ro,\yo'=\yo)$, 
and integrate over $\yo'$. This leads to a delta
function in $(\kk-\pp)$ and  Eq.~(\ref{corrlmaineq1}) becomes,
\bea
&&\hat{M}(\kk, t) =  
e^{-2\kappa\tau\kk^2}\int \bra{R} 
M(\ro,t_0)e^{-i\kk\cdot\ro}d^3\ro \nonumber \\ 
\label{magCorr}
&&\bra{R}
=\bbra{e^{-i\tau(\aaa\cdot\kk) (\sin{\bar A} - \sin{\bar B})}} 
\label{avterm}
\eea
where, $\bar A=(\xo\cdot\q+\psi)$ and $\bar B=(\yo\cdot\q+\psi)$.
Note that $\bra{R}$ is expected to be only a function
of $\ro$, due to statistical homogeneity of the renewing flow, 
as we also see explicitly in Sec.~\ref{GK}.

We are interested in the limit of large Peclet number, $\Rk\gg1$.
In this limit, we can  
expand the exponential in the diffusive Green's function
and consider only leading order term in $\kappa$. Then we have
\bea
\hat{M}(\kk, t) &=&  (1 -2\kappa\tau\kk^2) \int \bra{R} 
M(\ro,t_0)e^{-i\kk\cdot\ro}d^3\ro \nonumber \\ 
&=& \int \bra{R} \hat{M}(\pp,t_0) e^{i(\pp-\kk)\cdot\ro} d^3\ro
\frac{d^3\pp}{(2\pi)^3} - 2\kappa \tau \kk^2 \hat{M}(\kk,t_0)
\label{Fourspec}
\eea
In the diffusive term, we consider only the $\tau$ independent term
of $\bra{R}$
to multiply with $2\kappa\tau\kk^2$, as all the other terms will be 
of the order $\kappa\tau^2$ or higher. (Specifically, in the
independent $\kappa \to 0$ limit, we neglect 
terms proportional to $\kappa\tau^2$, compared to those 
$\propto \tau^3$ and $\tau^4$.)
We will use \Eq{Fourspec} when we generalize the Kraichnan
evolution equation for the passive scalar, to incorporate finite
$\tau$ effects.

We can also study the evolution of scalar fluctuations 
in terms of the real space correlation function. For this we 
take the inverse Fourier transform of $\hat{M}(\kk, t)$ and get, 
\bea
M(\rr, t) &=& \int  
(1-2\kappa\tau\kk^2) \bra{R} 
M(\ro,t_0)e^{i\kk\cdot(\rr-\ro)}d^3\ro \frac{d^3\kk}{(2\pi)^3} \nonumber \\
&=&\int \bbra{R} M(\ro, t_0) )e^{i\kk\cdot(\rr-\ro)}d^3\ro \frac{d^3\kk}{(2\pi)^3} 
+ 2\kappa\tau\nabla^2 M(\rr, t).  
\label{invFT}
\eea
Here we again expanded the exponential in the diffusive Green's function
relevant in the independent small $\kappa$ (or $\Rk \gg 1$) limit.
And in the diffusive term of \Eq{invFT} consider only the $\tau$ 
independent term in $\bra{R}$
as above.
Then we write $(-\kk^2)(e^{i\kk\cdot\rr})$ as 
$\nabla^2 e^{i\kk\cdot\rr}$ and can take $\nabla^2$ out of the integral.

Both the Fourier space and real space approaches are useful in
different contexts. We now turn to the evaluation of $\bra{R}$.

\section{The generalized Kraichnan equation}
\label{GK}

The exact analytical evaluation of $\bra{R}$ turns out to be difficult. 
However, for small 
renewal times $\tau$ or dimensionless Kubo number defined as
$K = q \vert\aaa\vert \tau = q a\tau$, 
we can motivate a perturbative,  
Taylor series expansion of the exponential in  $\bra{R}$, as follows.
 
Firstly the exponent in $\bra{R}$ contains the term 
$2(\sin{\bar A} - \sin{\bar B}) 
= \sin(\q\cdot\ro/2) \cos(\psi + \q\cdot\RR_0)$,
where $\RR_0 = (\xo+\yo)/2$. 
Also note that $\kk$ and $\ro$ are in general Fourier
conjugate variables and will satisfy $kr_0 \sim 1$, where
$k=\vert \kk \vert$ and $r_0 = \vert \ro\vert$.
We can consider two cases. If $qr_0 \ll 1$, then 
$\sin(\q\cdot\ro) \sim \q\cdot\ro$. 
Consequently the phase of the
exponential in Eq.~(\ref{avterm}) is of order 
$(k a \tau q r_0) \sim q a \tau = K$.
On the other hand when $qr_0 \sim 1$, 
$ka\tau \sim a\tau/r_0 \sim q a \tau =K$ again.
(note that for $qr_0\gg1$, modes with $k\sim1/r_0$ will
contribute to the correlator, and $ka\tau \sim a\tau/r_0 \ll 1$
again for small enough $\tau$).
Thus in all cases when
$K \ll 1$, one can expand the exponential
in Eq.~(\ref{avterm}) in powers of $\tau$. We do this
retaining terms up to $\tau^4$ order. On keeping up to 
$\tau^2$ terms in Eq.~(\ref{avterm}), we recover the Kraichnan 
evolution equation for the passive scalar, while the $\tau^4$ terms
give finite-$\tau$ corrections.
On expansion we have,
\be
\bra{R}=\bbra{\left[1 - i\tau\sigma - \frac{\tau^2\sigma^2}{2!} 
+ \frac{i\tau^3\sigma^3}{3!} + \frac{\tau^4\sigma^4}{4!}\right] }, 
\label{expterms}
\ee
where 
\be
\sigma=(\aaa\cdot\kk)(\sin{\bar A} - \sin{\bar B}) 
= 2(\aaa\cdot\kk) \sin\left(\frac{\q\cdot\ro}{2}\right)
\cos\left(\psi + \q\cdot{\bf R}_0\right)
\label{sigma}
\ee

\subsection{Kraichnan passive scalar equation from terms up to order $\tau^2$}

We first consider terms in \Eq{expterms} up to the order $\tau^2$
and average over $\psi$, $\hat{\bf a}$ and $\hat{\bf q}$.
To begin with we note that $\bra{\sigma}=0$ as it is proportional
to $\cos(\psi +...)$ and the integral over $\psi$ vanishes. In the
third term of \Eq{expterms}, we have 
\be
\sigma^2 =
4(\aaa\cdot\kk)^2 \sin^2\left(\frac{\q\cdot\ro}{2}\right)
\cos^2\left(\psi + \q\cdot{\bf R}_0\right)
=  (\aaa\cdot\kk)^2 (1- \cos\q\cdot\ro)
(1+\cos(2\psi + 2\q\cdot{\bf R}_0))
\label{sig2}
\ee
and on averaging over $\psi$, the term proportional
to $\cos(2\psi +..)$ vanishes. Therefore we have for this term
\be
\frac{\tau^2\bra{\sigma^2}}{2}
= \frac{\tau^2}{2}k_ik_j\bra{a_ia_j(1- \cos\q\cdot\ro)}
=2\tau k_ik_j[T_{ij}(0) - T_{ij}(r_0)].
\label{3rd}
\ee
where we have defined
the turbulent diffusion tensor, in terms of 
the two point velocity correlator, as
\be
T_{ij}(\ro) = \frac{\tau}{2}\bra{u_i(\xo)u_j(\yo)} 
=\frac{\tau}{2} \bra{a_i a_j \sin(\bar A) \sin(\bar B)}=
 \frac{\tau}{4}\bra{a_i a_j \cos(\q\cdot\ro)}.
\label{difften}
\ee
We have included a factor of $\tau$ in the
above definition of $T_{ij}$, 
since in the limit of $\tau\to0$, 
(as for delta-correlated in time flow) one still wants the
turbulent diffusion to have a finite limit.
Using \Eq{3rd}, we then have up to order $\tau^2$,
\be
\bra{R} = 1 - 2\tau k_ik_j[T_{ij}(0) - T_{ij}(r_0)].
\label{Rtau2}
\ee
We substitute \Eq{Rtau2} in \Eq{invFT} for the passive
scalar correlation function. For the term multiplying unity,
the integration in \Eq{invFT} over $\kk$ gives a delta 
function $\delta^3(\rr-\ro)$, and thus the integration
over $\ro$ simply replaces it with $\rr$ in $M(\ro,t_0)$.
For the second term containing $p_i$, we can first write
it as a derivative with respect to $r_i$, pull it out of the
integral, and then again integrate over $\kk$ and $\ro$ 
as above. We then get from \Eq{Rtau2} and \Eq{invFT},
\bea
&&M(r,t) = \int \left(1-2\tau k_ik_j[T_{ij}(0) - T_{ij}(r_0)]\right)
M(\ro,t_0)~e^{i\kk\cdot(\rr-\ro)}~d^3\ro \frac{d^3\kk}{(2\pi)^3} 
\nonumber \\
&& \hskip 3 true in + \ 2\kappa\tau\nabla^2 M(\rr, t)
\nonumber \\
&=& M(r,t_0) + 2\tau \partial_i\partial_j \int \left[T_{ij}(0) - T_{ij}(\ro)\right]
M(\ro,t_0)~e^{i\kk\cdot(\rr-\ro)}~d^3\ro \frac{d^3\kk}{(2\pi)^3}
\nonumber \\
&& \hskip 3 true in + \ 2\kappa\tau\nabla^2 M(\rr, t)
\nonumber \\
&=&M(r,t_0)+ 2\tau~\partial_i\partial_j\left[(T_{ij}(0)-T_{ij}(\rr))
M(r,t_0)\right]
+ 2\kappa\tau\nabla^2 M(r, t).
\label{mcor}
\eea
Here $\partial_i = \partial/\partial r_i$ and we have used
the fact that for statistically isotropic scalar correlations 
$M(\rr,t) = M(r,t)$.
Note that for a divergence free velocity, $\partial_i(T_{ij})=0$
and so the spatial derivatives can be taken through the $T_{ij}$ terms
\Eq{mcor} to directly act on $M(r,t)$.

A differential equation for the time evolution of $M(r,t)$ can be derived
from \Eq{mcor}, by dividing throughout by $\tau$, and noting that
in the limit $\tau \to 0$,
$(M(r,t) -  M(r,t_0))/\tau=\partial M/\partial t$. We get
\be
\frac{\partial M(r,t)}{\partial t} = 
\left[T_{ij}(0)-T_{ij}(\rr)\right]\frac{\partial^2M}{\partial r^i \partial r^j}
+ 2\kappa\nabla^2 M(r, t) 
\label{evol1}
\ee
This can be further simplified by first noting that
\be
\frac{\partial^2M}{\partial r^i \partial r^j}
= \left(\delta_{ij} -\rvu{i}\rvu{j}\right) 
\frac{1}{r}\frac{\partial M}{\partial r}
+\rvu{i}\rvu{j} \frac{\partial^2 M}{\partial r^2}.
\label{d2M}
\ee
Also for a statistically homogeneous, isotropic and non helical 
velocity field, the correlation function
\be
T_{ij} = \left(\delta_{ij} -\rvu{i}\rvu{j}\right) T_{\rm N}(r,t)
+\rvu{i}\rvu{j} T_{\rm L}(r,t).
\label{Tij}
\ee
Here $\hat{r_i}=r_i/r$, and $T_{ij}(0)=\delta_{ij}T_L(0)$, with
\be
T_{L}(r,t) =\rvu{i}\rvu{j} T_{ij} \quad {\rm and} \quad 
T_{N}(r,t) = \frac{1}{2r}\frac{\partial (r^2 T_L)}{\partial r}
\label{tltn}
\ee
respectively being, the longitudinal
and transversal correlation functions of the velocity field.
Using \Eqs{d2M}{Tij} in \Eq{evol1}, we then get
\be 
\frac{\partial M(r,t)}{\partial t} = 
\frac{2}{r^2}\frac{\partial}{\partial r} 
\left[ r^2 \left(\kappa+T_L(0) - T_L(r)\right) \frac{\partial M}{\partial r}\right]
\label{Kraich1}
\ee
This is the Kraichnan equation for the evolution of the passive
scalar fluctuations in real-space. It has a very simple interpretation;
at the lowest order in $\tau$ the effect of the random velocity
is to add scale dependent turbulent diffusivity $[T_L(0)-T_L(r)]$
to the microscopic diffusivity $\kappa$.
Although we derived \Eq{Kraich1} 
in the context of renewing flows, the only reflection
of the velocity field is in $T_{ij}$. In fact the
same equation obtains when one uses a delta correlated in time
velocity field with the spatial part of the correlation being given
by $T_{ij}$ (see also \citep{Zeld88b}).

\subsection{Kraichnan equation in Fourier space from up to $\tau^2$ terms}

In order to derive the 
passive scalar
spectrum 
(both for the steady state and the decaying case),
in a transparent manner, 
it is more useful to directly work in Fourier space. 
In Fourier space, \Eq{Kraich1} becomes an integro-differential
equation. However, the Batchelor 
regime is supposed to obtain only for wavenumber $k$, much larger than the
viscous cut-off wavenumber (in the case of single scale renewing flow
this corresponds to $q$) and much smaller than the wavenumber where
scalar diffusion is important; or in the range $q\ll k\ll k_\kappa$. 
For $k\gg q$, one has $qr\ll 1$ and so we can make the small $qr$ approximation
in evaluating $T_{ij}(r)$.

For this we expand $\cos(\q\cdot\rr)$ in \Eq{difften} for small $qr$,
use \Eq{aEns} to write $a_i$ in terms of $A_i$.
We also use  $\bra{A_iA_j} =A^2 \delta_{ij}/3$,
$\bra{q_iq_j} = q^2 \delta_{ij}/3$ and
$\bra{q_iq_jq_lq_m} = (4/15)[\delta_{ij}\delta_{lm}
+ \delta_{il}\delta_{jm}+\delta_{im}\delta_{lj}]$ to get
$T_{ij}(0) = \delta_{ij} T_L(0)$, with $T_L(0) = \kappa_t = A^2\tau/18$ and
\be
[T_{ij}(0)-T_{ij}(\rr)] = \frac{\tau}{4}\bra{a_ia_j[1-\cos(\q\cdot\rr)]}
= {\cal A} \left[2 r^2 \delta_{ij} - r_ir_j\right]
\label{ToTr}
\ee
with ${\cal A} = A^2q^2\tau/90 = \kappa_t q^2/5$.
From \Eq{ToTr} we then have for small $qr$, and up to order $\tau^2$,
\be
\bra{R} = 1 - \tau k_i k_j {\cal A}(2r_0^2\delta_{ij} - r_{0i}r_{0j})
\equiv 1-R_2
\label{R2}
\ee
We use this in \Eq{Fourspec} to derive the Fourier space Kraichnan equation 
and passive scalar spectrum in the viscous-convective regime. 

We substitute \Eq{R2} in \Eq{Fourspec} for the passive scalar spectrum,
in the limit $k \gg q$, and up to order $\tau^2$,
\be
\hat{M}(\kk, t) 
= \int (1 - R_2) \hat{M}(\pp,t_0) e^{i(\pp-\kk)\cdot\ro} d^3\ro
\frac{d^3\pp}{(2\pi)^3} 
- 2\kappa \tau \kk^2 \hat{M}(\kk,t_0)
\label{FourspecK}
\ee
In the first term proportional to unity one can straightaway do the  
integration over $\ro$ to get a delta-function $\delta^3(\pp-\kk)$ and
then the $\pp$ integral becomes trivial and gives $\hat{M}(\kk,t_0)$. 
In the second term proportional to $R_2$,
we convert each factor or $r_{0i}$ to a derivative with respect to $k_i$ and
integrate by parts. The integral over $\ro$ again gives $\delta^3(\pp-\kk)$
and again the integral over $\pp$ can be done trivially.
Then 
\bea
-\int R_2 \hat{M}(\pp,t_0) e^{i(\pp-\kk)\cdot\ro} d^3\ro
\frac{d^3\pp}{(2\pi)^3} 
&=& {\cal A}\tau k_ik_j \left[2\delta_{ij}  
\frac{\partial^2 \hat{M}}{\partial k_l\partial k_l}
-\frac{\partial^2 \hat{M}}{\partial k_i\partial k_j} \right]
\nonumber \\
&=& {\cal A}\tau \left[k^2  
\frac{\partial^2 \hat{M}}{\partial k^2}
+4k\frac{\partial \hat{M}}{\partial k} \right].
\label{R2value}
\eea
Substituting \Eq{R2value} into \Eq{FourspecK}, dividing throughout
by $\tau$ and taking the $\tau\to 0$ limit we then get,
\be
\frac{\partial \hat{M}(k,t)}{\partial t}
= \frac{\kappa_t q^2}{5}\left[k^2  
\frac{\partial^2 \hat{M}}{\partial k^2}
+4k\frac{\partial \hat{M}}{\partial k} \right]
- 2\kappa k^2 \hat{M}(k,t)
\label{kraichp}
\ee
This is \citet{K68} evolution equation for the passive scalar in
Fourier space applicable to the viscous-convective regime, with
the first term on the right hand side of \Eq{kraichp} the same
as $T(k,t)/4\pi k^2$ given in Eq. (3.6) of K68.
We can also rewrite \Eq{kraichp} in terms of an equation for
the 1-D scalar spectrum $E_\theta(k) = 4\pi k^2 \tilde{M}(k)$.
We get,
\be
\frac{\partial E_\theta(k,t)}{\partial t}
= \frac{\kappa_t q^2}{5}\frac{\partial}{\partial k}\left[k^4  
\frac{\partial }{\partial k}\left(\frac{E_\theta}{k^2}\right)\right]
- 2\kappa k^2 E_\theta(k,t).
\label{kraich1d}
\ee
We look at its consequences for the passive scalar spectrum 
after also working out the finite $\tau$ corrections to \Eq{kraichp}.

\subsection{Finite $\tau$ correction to Kraichnan equation from 
up to $\tau^4$ terms}

Now let us consider the higher order terms up to $\tau^4$.
Firstly, the $\tau^3$ term in \Eq{expterms}, is proportional to 
$\bra{\cos^3(\psi +\q\cdot{\bf R}_0)}$, always contain a $n\psi$
in the argument of a cosine. Thus it goes to zero on averaging
over $\psi$. Now consider the term in \Eq{expterms} of order $\tau^4$.
This is given by
\bea
R_4 = \frac{\bra{\tau^4\sigma^4}}{24}
&=&\frac{\tau^4}{24}
\bra{(\aaa\cdot\kk)^4 \sin^4\left(\frac{\q\cdot\ro}{2}\right)
\cos^4\left(\psi + \q\cdot{\bf R}_0\right)}
\nonumber \\
&=&\frac{\tau^4}{16}k_m k_n k_r k_s
\bbra{a_m a_n a_r a_s\left(\frac{3}{2}-2\cos(\q\cdot\ro)
+\frac{\cos(2\q\cdot\ro)}{2}\right)} \nonumber \\
&=&\tau^2 k_m k_n k_r k_s
\left[\frac{T_{mnrs}^{x^2y^2}}{4}
-\frac{T_{mnrs}^{x^3y}}{3}+\frac{T_{mnrs}^{x^4}}{12}\right],
\label{tau4}
\eea
where we have carried out the averaging over $\psi$.
Further, in the last line of \Eq{tau4}, we have
expressed $R_4$ in terms of 
the two point fourth order velocity correlators as defined
by BS15. Note that
three such velocity correlators can be defined,
\bea
\label{Tmnrs1}
T_{mnrs}^{x^2y^2}&=&\tau^2\bra{u_m(\x)u_n(\y)u_r(\x)u_s(\y)} 
=\frac{\tau^2}{4}\bbra{a_m a_n a_r a_s 
\left(1+\frac{\cos(2\q\cdot\ro)}{2}\right)},
 \\
\label{Tmnrs2}
T_{mnrs}^{x^3y}&=&\tau^2\bra{u_m(\x)u_n(\x)u_r(\x)u_s(\y)} 
=\frac{3\tau^2}{8}\bbra{a_m a_n a_r a_s \cos(\q\cdot\ro)},
 \\
\label{Tmnrs3}
T_{mnrs}^{x^4}&=&\tau^2\bra{u_m(\x)u_n(\x)u_r(\x)u_s(\x)}
=\frac{3\tau^2}{8}\bbra{a_m a_n a_r a_s }.
\eea
Where in the last parts of each equation, the $\psi$
averaged expressions are given. Further, we have also multiplied
the fourth order velocity correlators
by $\tau^2$, as we envisage that 
$T_{ijkl}$ 
will be finite even in the $\tau\to 0$ limit. Essentially, they
behave like products of turbulent diffusion.
Interestingly, the renewing flow is not Gaussian random,
and therefore higher order correlators of $\uu$ are not
just the product of two-point correlators.

We can now add the $R_4$ contribution to $\bra{R}$, substitute
in to \Eq{invFT} for the scalar correlator. Again each $k_i$ can be
converted to a derivative over $r_i$ and pulled out of the integral in
\Eq{invFT}. The resulting integrals over $\kk$ and $\ro$,  
including the $R_4$ term can then  be carried out as earlier in
deriving \Eq{mcor}. We then divide all contributions by $\tau$
and take the limit as $\tau\to 0$. For the $\tau^4$ term
on division by $\tau$, a factor of $\tau$ remains as a
small effective finite time parameter. We then get for
the extended Kraichnan evolution equation, 
\bea
\frac{\partial M(r,t)}{\partial t} &=& 
\left[T_{ij}(0)-T_{ij}(\rr)\right]\frac{\partial^2M}{\partial r^i \partial r^j}
+ 2\kappa\tau\nabla^2 M(r, t) \nonumber \\ 
&+& \tau\left[\frac{T_{mnrs}^{x^2y^2}}{4} - \frac{T_{mnrs}^{x^3y}}{3}
+ \frac{T_{mnrs}^{x^4}}{12}\right]M_{,mnrs}  
\label{finalscalcor}
\eea
The first line in \Eq{finalscalcor} have the terms which give 
the standard Kraichnan equation \Eq{evol1} in real space, whereas  
the second line contains the finite correlation time corrections.
This generalized Kraichnan equation can also be further simplified by
using an explicit form for the fourth order correlators and the
fourth derivative of $M$, following very similar algebra as in BS15. 
It can then be used to examine the finite correlation time modification to the passive scalar
spectrum both in the steady state and decaying cases.
However this is again derived in a more transparent manner in Fourier space,
to which we now turn.

\subsection{Generalized Kraichnan equation in Fourier space}
  
For calculating the finite $\tau$ corrections to the Fourier
space K68 equation \Eq{kraichp} and the 
passive scalar spectrum in the Batchelor regime,
we need only the small $qr_0$ expansion to the $R_4$ term.
This is because, as we discussed earlier,  an extended
Batchelor regime obtains for large Schmidt numbers, 
in the range of wavenumbers when $k_\kappa\gg k\gg q$, 
From \Eq{tau4}, in the $qr_0\ll 1$ limit we have
\be
R_4 = \frac{\tau^4}{64}
\bra{(\aaa\cdot\kk)^4(\q\cdot\ro)^4}.
\label{R4smallr}
\ee
Substituting \Eq{R4smallr} into \Eq{Fourspec}, the additional $\tau^4$
contribution of $R_4$ to $M(\kk,t)$ is given by
\bea
\hat{M}_4(k, t) &=&  
\frac{\tau^4}{64}
\bra{(\aaa\cdot\kk)^4(q_aq_bq_cq_d)} \int r_{0a} r_{0b} r_{0c} r_{0d}
\hat{M}(\pp,t_0) e^{i(\pp-\kk)\cdot\ro} d^3\ro
\frac{d^3\pp}{(2\pi)^3} \nonumber \\ 
&=& 
\frac{\tau^4}{64}
\bra{(\aaa\cdot\kk)^4(q_aq_bq_cq_d)}\frac{\partial^4 \hat{M}(k,t_0)}
{\partial k_a \partial k_b \partial k_c \partial k_d}.
\label{R4MK}
\eea
In the second step of \Eq{R4MK}, we have converted each $r_{0a}$ factor to 
a derivative of the exponential factor with respect to $p_a$ and
integrated by parts to transfer these to derivatives of $\hat{M}$.
The integral over $\ro$ then gives a delta function $\delta^3(\kk-\pp)$
and the integral over $\pp$ then simply replaces $\pp$ everywhere
by $\kk$.

We can again write for each 
$\aaa\cdot\kk= a_ik_i = [\delta_{ij} -\hat{q}_i\hat{q}_j]A_jk_i$,
and average over the independent $A_i$. This leaves averaging over $\q$. 
If one does this $q_i$ averaging
naively, we would need to deal with averaging an eight-point $q_i$
correlator, which involves 105 terms. To avoid this we first
evaluate the fourth derivative of $\hat{M}$ explicitly, which gives 
\bea
\frac{\partial^4 \hat{M}}
{\partial k_a \partial k_b \partial k_c \partial k_d}
 &=& \frac{1}{k}\frac{\partial}{\partial k}
\left(\frac{1}{k}\frac{\partial M}{\partial k}\right) \delta_{(ab}\delta_{cd)}
+ \frac{1}{k}\frac{\partial}{\partial k}\left[
\frac{1}{k}\frac{\partial}{\partial k}
\left(\frac{1}{k}\frac{\partial M}{\partial k}\right)\right] 
\delta_{(ab}k_ck_{d)}
\nonumber \\
&+& \frac{1}{k}\frac{\partial}{\partial k}\left(
\frac{1}{k}\frac{\partial}{\partial k}\left[
\frac{1}{k}\frac{\partial}{\partial k}
\left(\frac{1}{k}\frac{\partial M}{\partial k}\right)\right]\right)
k_ak_bk_ck_d.
\label{Mabcd}
\eea
We have used the brackets $()$ in the subscripts
of two second rank tensors, to denote 
addition of all terms from different permutations 
of the four indices considered in pairs. 
This brings out components of $k_a$, which can combine with corresponding
$q_a$'s, and one only gets up to eighth power of the dot product
$(\q\cdot\kk)= qk\mu$, where $\mu$ is uniformly random over the interval 
$(-1,1)$. Thus one only has to do integrations of the form
$\int \mu^n d\mu$ for the $q_i$ averaging, which become trivial.
Carrying out the above steps, a lengthy but straightforward calculation gives,
\be
\frac{\hat{M}_4(k, t)}{\tau} =
\frac{9}{350}\tau (\kappa_t q^2)^2
\left[ k^4 \frac{\partial^4 \hat{M}}{\partial k^4}   
+ 12 k^3 \frac{\partial^3 \hat{M}}{\partial k^3}   
+24 k^2 \frac{\partial^2 \hat{M}}{\partial k^2}   
-24 k \frac{\partial \hat{M}}{\partial k}   
\right]
\label{M4tau}
\ee
We define a dimensionless time co-ordinate $\bar t = t\kappa_t q^2$ and
$\bar\tau = \tau \kappa_t q^2$, dividing by a turbulent diffusion timescale
$(\kappa_t q^2)^{-1}$. Again taking the limit $\tau\to 0$, keeping
$\kappa_t$ fixed, and adding in the contribution due to $\hat{M}_4$,
we get for the generalized Kraichnan equation
\bea
\frac{\partial \hat{M}}{\partial \bar{t}}
&=& \frac{1}{5}\left[k^2  
\frac{\partial^2 \hat{M}}{\partial k^2}
+4k\frac{\partial \hat{M}}{\partial k} \right]
- 2\frac{\kappa}{\kappa_t} \frac{k^2}{q^2} \hat{M} \nonumber \\
&+& \bar\tau \frac{9}{350}
\left[ k^4 \frac{\partial^4 \hat{M}}{\partial k^4}   
+ 12 k^3 \frac{\partial^3 \hat{M}}{\partial k^3}   
+24 k^2 \frac{\partial^2 \hat{M}}{\partial k^2}   
-24 k \frac{\partial \hat{M}}{\partial k} \right]  
\label{kraichfinal}
\eea
Again the first line of \Eq{kraichfinal} is the standard K68 equation
in Fourier space, while the second line gives the leading order
finite $\tau$ correction to the Kraichnan equation. 
The generalized Kraichnan equation allows eigen-solutions 
of the form $\hat{M}(k,t) = \tilde{M}(k) e^{-\gamma \bar{t}}$.
We consider such solutions in the next section. Note that to get a steady state solution with $\gamma=0$, 
one would also need a source term at small $k$ as discussed 
earlier; if there is no source the scalar fluctuations would decay.
We consider both the steady state and decaying cases below. 
The decaying case, which is more subtle, needs 
a different treatment, obtained 
by solving for the evolution of $\hat{M}$ as an initial value
problem in terms of the Green function solution of \Eq{kraichfinal}.
The eigen solutions are useful in providing 
basis eigenmodes and eigenvalues for deriving the Green's function,
and thus we keep $\gamma$ to be non-zero below.

\section{The passive scalar spectrum at finite correlation time}

For determining the passive scalar spectrum at finite $\tau$
in the viscous-convective scales, we analyze \Eq{kraichfinal}
in more detail.
This evolution equation 
also has higher order (third and fourth) $k$-space derivatives  
when going to finite-$\tau$ case.
However as in BS15, these higher
derivative terms only appear as perturbative terms 
multiplied by the small parameter $\bar\tau$. 
It is then possible to follow the methodology of BS15 and use 
the Landau-Lifshitz type approximation,
earlier used in treating the effect of radiation reaction
force in electrodynamics (see \citet{LL} section 75).
In this approximation, the perturbative terms proportional
to $\bar\tau$ are first neglected, which gives basically Kraichnan 
equation for the passive scalar spectrum  
$\tilde{M}(k)$, and uses this 
to express higher order derivatives, $(\partial^3\tilde{M}/\partial k^3)$ 
$(\partial^4\tilde{M}/\partial k^4)$ 
in terms of the lower order $k$-derivatives,
$(\partial^2\tilde{M}/\partial k^2)$ 
and $(\partial\tilde{M}/\partial k)$. 

Further, 
for high $R_\kappa \gg 1$ and 
in the range of wavenumbers where 
the Batchelor regime obtains
we can also neglect the diffusive terms
in determining the higher order derivatives. In this limit, \Eq{kraichfinal} gives at the zeroth order in $\bar{\tau}$,
\be
k^2 \tilde{M}'' = -4 k\tilde{M}' + 5\gamma_0 \tilde{M},
\label{zerothk}
\ee
where a prime denotes a $k$-derivative $d/dk$ and
we have denoted by $\gamma_0$ the 
eigenvalue which obtains for the
Kraichnan equation in the $\tau\to 0$ limit.
We differentiate this expression first once and then twice to get,
\be 
 k^3\tilde{M}''' 
=  -4 k^2 \tilde{M}''
+ (4-5\gamma_0)\tilde{M}'
+10 \gamma_0 \tilde{M},
\label{m3}
\ee
\be
k^4 \tilde{M}''''
= (24-5\gamma_0)k^2 \tilde{M}'' 
+(40\gamma_0 -24) k \tilde{M}'
-70 \gamma_0 \tilde{M}. 
\label{m4}
\ee
Substituting now \Eqs{m3}{m4} back into 
\Eq{kraichfinal}, the generalized Kraichnan equation, 
to the leading order in $\bar\tau$, then becomes
\bea
k^2 \tilde{M}''\left(\frac{1}{5}-\frac{9\gamma_0\bar\tau}{70}\right)
+k\tilde{M}'\left(\frac{4}{5}-\frac{36\gamma_0\bar\tau}{70}\right)
+\left(\gamma+\frac{9}{7} \gamma_0\bar\tau \right)\tilde{M}
-2 \frac{\kappa}{\kappa_t} \frac{k^2}{q^2} \tilde{M} =0
\label{kraichLL}
\eea
Remarkably, the coefficients in the generalized Kraichnan equation 
\Eq{kraichfinal} are such that all perturbative terms which do not depend on
$\gamma_0$ cancel out in \Eq{kraichLL}. Thus for steady state 
$\gamma_0=0=\gamma$, the Kraichnan equation for $\tilde{M}(k)$, 
for a finite correlation time $\tau$, remains the same as that 
derived under the delta-correlation in time approximation. 
This also means that the Batchelor spectrum is unchanged for
a finite $\tau$ to leading order in $\tau$.
We will also show below the equally important result
that the passive scalar spectrum for the decaying case is also unchanged
for finite $\tau$ to leading order.
To see these results explicitly, we present below both a scaling solution
to the eigenfunction, valid in the viscous-convective regime, and also a more exact treatment
which takes into account the diffusive term.

\subsection{The scalar spectrum from a scaling solution}

Consider first the scalar spectrum
in the viscous-convective regime, where $q \ll k\ll k_\kappa$.
Thus \Eq{kraichfinal} or \Eq{kraichLL} are applicable 
and the diffusive term can also be neglected.
In this regime we see that \Eq{kraichLL}, and in fact the original 
\Eq{kraichfinal}, are scale free, as scaling $k \to  ck$ leaves
them invariant. Therefore they admit power law solutions, 
of the form $\tilde{M}(k) = M_0 k^{-\lambda}$, where from \Eq{kraichLL},
$\lambda$ is determined by
\be
\lambda^2 -3\lambda +\frac{5[\gamma+(9/7)\gamma_0\bar\tau]}
{1- (9/14) \gamma_0\bar\tau} =0; \quad
\lambda =\frac{3}{2} \pm D; \quad 
D= \frac{3}{2} \left[1 - \frac{(20/9)[\gamma+(9/7)\gamma_0\bar\tau]}
{(1- (9/14) \gamma_0\bar\tau)}\right]^{1/2}. 
\label{lambda}
\ee 
One can consider two cases:
\subsubsection{The steady state solution}
\label{steady2}

For a steady state situation, this reduces simply to $\lambda^2-3\lambda=0$
independent of $\tau$, whose solutions are $\lambda=3$ and $\lambda=0$.
(In this case $D=3/2$).
The case $\lambda=0$ corresponds to a case where there is 
no transfer of scalar variance from scale to scale \citep{K68}. 
Interestingly, \citet{NL2000} point out that this case corresponds to
a constant flux of what they term as pseudo energy, 
whose 1-D spectrum is given by $\tilde{M}(k)/k$.
The case which is relevant for us is when $\lambda=3$, where the
rate of transfer of scalar variance from wavenumbers smaller than $k$
to those larger is constant (see K68).
(To regularize the small $k$ behaviour of this solution requires
there be a source which changes the power law behaviour at some small $k_0$
as discussed earlier).
Note that the 1-d scalar spectrum, say $E_\theta(k) = 4\pi k^2 \tilde{M}(k)$
and therefore for $\lambda=3$, we have 
$E_\theta(k) \propto 1/k$, or the Batchelor spectrum.
Therefore, we obtain the Batchelor spectrum as the steady
state solution, in the viscous-convective range, 
of even the generalized Kraichnan equation that takes into account,
the leading order effects of a finite $\tau$.
Such a result has also been obtained earlier in the perturbative limit
and weak non-Gaussianity by \citet{CFL96}. And for arbitrary $\tau$
using a Lagrangian analysis with simplifying assumptions \citep{FGV01,FL03}, 
as discussed also  in Appendix~\ref{appA}. Our approach above is complementary 
and has used a generalization of the Kraichnan differential equation which is
applicable even to the strongly non-Gaussian renewing flow.

\subsubsection{The case of finite eigenvalue $\gamma$}

Now turn to the case when there is no source of scalar
fluctuations, and so they necessarily decay with,
possibly a superposition of eigenfunctions with finite $\gamma\ne 0$.
An approximate value for the leading (smallest) eigenvalue $\gamma$, for $R_\kappa \gg1$, can be got
following an argument from \citet{GCS96}, who
looked at a similar power law solution of the small scale
dynamo problem. 
In order to satisfy the boundary condition
that the $k^{-\lambda}$ eigenmode is regular at both large and small $k$, they argued that the limiting eigenvalue $\gamma$
(or the growth rate in the dynamo problem) is given by $\gamma(\lambda_m)$, 
with $\lambda_m$ the value of $\lambda$ where $d\gamma/d\lambda=0$. This leads to
\be
\gamma_0 \approx  \frac{9}{20} =\bar{\gamma}_0, \quad {\rm and}
\quad
\gamma \approx  \frac{9}{20}\left(1 -\frac{63}{40}\bar\tau \right)
=\bar{\gamma}.
\label{gammascal}
\ee
Note that \Eq{gammascal} also implies that $D\approx0$. We will
see below that, a more exact treatment picks out a small imaginary value of 
$D$, so as to also satisfy the regularity condition on $\tilde{M}(k)$ at small $k$.
The fact that $D$ is imaginary and small implies
that the two roots of $\lambda$ in \Eq{lambda} are
complex conjugate, with a small imaginary value of
$\lambda$. More importantly, the real part is $\lambda_R =3/2$
independent of $\tau$, and independent of the value of the eigenvalue $\gamma$.
The eigenfunctions (from the power law solution, 
$k^{-3/2 \pm i\vert D\vert}=
k^{-3/2}\exp(\pm i\vert D\vert\ln{k})$) are then of the form 
\be
\tilde{M}(k)= k^{-3/2} \sin(\vert D\vert \ln(k) + c_1),
\label{scalsoln}
\ee
where the only $\gamma$ and $\tau$ dependencies are in $D$ and the 
constant $c_1$ which are to be fixed from the boundary conditions
at large and small $k$. However, due to the expected small value of $D$ and the fact that the phase of the sine function is only logarithmically
dependent of $k$, we expect the dominant behaviour of the eigenfunctions to be independent of $\tau$.
This implies that the spectrum,
which would be superposition of such eigenfunctions (with different $\gamma$ or $D$) will also be independent of $\tau$.
This is an important new result, that the passive scalar spectrum in the decaying case is also independent of $\tau$ to leading
order in $\tau$. Further, one may also expect that the spectrum itself dominantly falls off as $\tilde{M}(k) \propto k^{-3/2}$ or $E_\theta\propto k^{1/2}$,
independent of $\tau$. We will check this in more detail below, when solving the initial value problem for the decay.
\footnote{
We note in passing that for the corresponding two-dimensional problem for passive scalar decay in the $\tau\to0$ limit,
\citet{NL2000} obtained a flat $k^0$ spectrum in the Batchelor regime. 
Both the two and three dimensional results are consistent
with the general expectation that, for the Kraichnan model 
in the Batchelor regime, the scalar eigenfunction during decay goes as $E_\theta \propto k^{d/2-1}$ 
in d-dimensions \citep{CL2003,SHC04}.
}
Further, from \Eq{gammascal}, we see that 
the the limiting decay rate $\gamma$ is of order of the turbulent
diffusion rate at the forcing scale, since we normalised $t$ by $(\kappa_tq^2)^{-1}$.
We also see that a finite $\tau$ decreases this rate at which the passive scalar decays, and thus the Kraichnan model over estimates
the decay rate. In the scaling solution, we neglected the diffusive term.
We now consider the effect of including this term.

\subsection{The scalar spectrum including diffusive effects}

In fact one can get an exact solution of \Eq{kraichLL} which
includes the diffusive term.  For this let us substitute 
$\tilde{M}(k) = k^{-3/2} K(k)$ into \Eq{kraichLL}.
This leads to a Modified Bessel equation for $K(k)$ given by
\be
x^2 \frac{d^2K}{dx^2} + x \frac{dK}{dx} -[x^2 +\nu^2]K = 0,
\label{bessel}
\ee
where we have defined $\nu^2\equiv D^2$, and 
\be
x = \frac{k}{k_d}, \quad 
k_d = q \left(\frac{\kappa_t}{10\kappa}\right)^{1/2},
\quad
\alpha = 1 - \frac{9}{14}\gamma_0\bar\tau, 
\quad
\nu^2 = \frac{9}{4\alpha} 
\left[ 1 - \frac{20}{9} \gamma - \frac{7}{2} \gamma_0\bar\tau \right].
\label{defn}
\ee
We note that the Peclet number $R_\kappa \sim \kappa_t/\kappa$, and 
so $k_d \sim q R_\kappa^{1/2}=k_\kappa$, the diffusive wavenumber 
and $x \sim k/k_d$ is basically $k$ normalised by the diffusive
wavenumber. 
Thus in the viscous-convective regime $x \ll 1$, while
in the diffusive regime, $x > 1$. Further, consistent with neglecting the effect of $\bar\tau$ on the diffusive
term in the extended Kraichnan equation, for $R_\kappa \gg 1$, 
we have also neglected its effect on $k_d$ 
above. (We note in passing that even at $k=q$, $x \sim 1/R_\kappa^{1/2} \ll 1$, for
large Peclet number $R_\kappa \gg 1$, although the small $qr$ approximation
will become inaccurate).  

\subsubsection{Steady state solution}

First consider the steady state solution with $\gamma=0=\gamma_0$.
As we remarked earlier, this assumes that there is a source of
scalar fluctuations which contributes at low wavenumbers, outside
the viscous convective range, so that \Eq{kraichLL} or \Eq{bessel}
are still valid. For the steady state case, we have $\nu^2=9/4$.
Then the two independent solutions of \Eq{bessel} are the modified 
Bessel functions of the first kind, $I_{3/2}(x)$ and of the second 
kind $K_{3/2}(x)$.
We also have to satisfy the boundary condition that 
as $x \to \infty$, $K(x) \to 0$. This rules out the solution
proportional to $I_{3/2}$, and so we have
\be
\tilde{M}(k) = k^{-3/2} K_{3/2}(k/k_d) 
\label{steady}
\ee
Note that as $x\to \infty$, $K_{3/2}(x) \to x^{-1/2} e^{-x}$ and so
$\tilde{M}(k) \to k^{-2} \exp(-k/k_d)$. This then implies that
$E_\theta(k) = 4\pi k^2 \tilde{M}(k) \to  \exp(-k/k_d)$, and so
dies exponentially beyond the diffusive wavenumber. 
Such an exponential fall off for the steady state case 
has been shown earlier by K68 and \citet{K74}.
And in the viscous-convective regime when $x \ll 1$, we have 
$K_{3/2}(x) \to \Gamma(3/2)2^{1/2} x^{-3/2}$, so
$\tilde{M}(k) \propto k^{-3}$ which implies $E_\theta(k) \propto 1/k$,
or the Batchelor spectrum. Thus as in the
the scaling solution above, the more complete solution in \Eq{bessel}
also shows that the Batchelor spectrum obtains
in the viscous-convective range $q \ll k \ll k_\kappa$, even for
a finite $\tau$, to the leading order in $\tau$. 
In addition \Eq{steady} extrapolates the Batchelor
spectrum smoothly to an exponential fall off in the diffusive regime,
in agreement with K68 and \citet{K74}.

\subsection{The scalar spectrum during decay}
\label{scaldecay}

Now consider the case when there is no source for scalar fluctuations.
We then expect that the passive scalar fluctuations will decay. We are interested in how the decay rate 
and the scalar power spectrum in this case
are altered due to the effect of a finite $\tau$.
For the Kraichnan problem  itself, where one considers 
passive scalar decay with $\bar\tau \to 0$,
a number of different approaches have been adopted, either by looking at the
decaying eigenfunction \citep{KA92,SHC04} or in terms of
solving \Eq{kraichfinal} as an initial value problem \citep{BF99,CL2003}.
Here we solve for the evolution of $\hat{M}$ as an initial value
problem in terms of the Green's function solution of \Eq{kraichfinal}
in the Landau-Lifschitz approximation.

For finding this Green's function, we seek eigenfunction solutions
to \Eq{bessel} which vanish at $x\to \infty$ and are regular as $x\to 0$. Such eigensolutions need to have $\nu^2=-(\bar\nu)^2 < 0$,
and are given by the modified Bessel function of imaginary order $\bar\nu >0$.
That is
\be
\tilde{M}(k) = \tilde{M}_{\bar\nu}(k) 
= k^{-3/2} K_{i\bar\nu}(k/k_d),
\label{decay}
\ee
where from \Eq{defn}, $\bar\nu$ is related to the eigenvalue $\gamma$,
\be
\bar\nu^2 = \frac{9}{4\alpha} 
\left[\frac{20}{9} \gamma + \frac{7}{2} \gamma_0\bar\tau -1 \right]
= \frac{9}{4} 
\frac{\left[20\gamma/9 + 7\gamma_0\bar\tau/2 -1 \right]}
{(1 - 9\gamma_0\bar\tau/14)}.
\label{barnu}
\ee
In the limit $x\to\infty$, 
$\tilde{K}_{i\bar\nu}(x) \to x^{-1/2} e^{-x}$ still and
the eigenfunction and hence the
spectrum vanishes for $k \gg k_d$, again falling off exponentially with $k$.
In the opposite limit of $x \to 0$, we have \citep{abra} (see also Chapter 10. by Olver and Maximon in \citep{NIST:DLMF})
\be 
\tilde{M}_{\bar\nu}(k) \to -\left(\frac{\pi}{\bar\nu\sinh(\pi\bar\nu)}\right)^{1/2}
k^{-3/2}\sin\left(\bar\nu \ln(k/2k_d) + c_{\bar\nu} \right). 
\label{decaysmallx}
\ee
Here $c_{\bar\nu}$ is a $\bar\nu$ dependent real constant defined by the relation,
\be
\Gamma(1+i\bar\nu) = \left(\frac{\pi\bar\nu}{\sinh(\pi\bar\nu)}\right)^{1/2}
\exp(-ic_{\bar\nu}),
\label{cnudef}
\ee
and so $c_{\bar\nu} \to \bar{c}\bar\nu $ for $\bar\nu \ll 1$, where $\bar{c}$ 
is the Euler constant. 
Expectedly, the solution $\tilde{M}$ in 
small $k/k_d$ limit, given in \Eq{decaysmallx} is the same as
the scaling solution given by \Eq{scalsoln}, with now the constants in that
solution fixed by matching also to the diffusive range.

It is interesting to note  that the differential operator, say ${\cal L}$, 
defined on the right hand side of the generalized Kraichnan equation 
\Eq{kraichfinal} is self adjoint with respect to the inner product
\footnote{We thank a referee for bringing this to our attention.}
\be
\langle{\tilde{M}_1(k),\tilde{M}_2(k)}\rangle 
= \int_0^\infty k^2 dk \tilde{M}_1^*(k) \tilde{M}_2(k).
\label{selfad}
\ee
The eigenfunctions of ${\cal L}$ are given by \Eq{decay} (in the Landau-Lifschitz approximation), with a real eigenvalue $-\gamma$,
where $\gamma$ is related to $\bar\nu^2=-\nu^2$ as in \Eq{barnu}. They can then be labled by the value of $\bar\nu$, and 
under the inner product \Eq{selfad}, also satisfy the orthogonality condition,
\be
\langle{\tilde{M}_{\bar\nu},\tilde{M}_{\bar\nu'}}\rangle 
= \int_0^\infty  \frac{dk}{k} \ K_{ i \bar\nu}(k/k_d) K_{i \bar\nu'}(k/k_d)
= \delta (\bar\nu - \bar\nu')  \left[\frac{\pi^2}{2\bar\nu \sinh(\pi\bar\nu)}
\right].
\label{ortho}
\ee
The latter orthogonality relation of the modified Bessel functions of
imaginary order is discussed for example in \citet{SB10}.
Thus the eigenfunctions $\tilde{M}_{\bar\nu}(k)$ given in \Eq{decay} are orthogonal under \Eq{selfad}
and also delta-function normalizable, like the free particle eigenfunctions in quantum mechanics.
This allows us to expand the scalar spectrum $\hat{M}(k,t)$
in terms of the orthogonal eigenfunctions of ${\cal L}$ as,
\be
\hat{M}(k,\bar{t}) = \int_0^\infty d\mu \ g(\mu,\bar{t}) \ \tilde{M}_{\mu}
= \int_0^\infty d\mu \ g(\mu,\bar{t}) \  k^{-3/2} K_{i\mu}(k/k_d).
\label{KLtrans}
\ee
(The above transform without the $k^{-3/2}$ factor is called the 
Kontorovich-Lebedev transform \citep{NIST:DLMF}).
This transform can be inverted by multiplying \Eq{KLtrans} by 
$\tilde{M}_{\mu'}(k)$, integrating over $k^2 dk$ and using the orthogonality
condition \Eq{ortho}, to get
\be
g(\mu,\bar{t}) = \frac{2\mu \sinh(\pi\mu)}{\pi^2}
\int_0^\infty dk \ k^2 \ \tilde{M}_{\mu}(k) \hat{M}(k,\bar{t}).
\label{invtrans}
\ee
We use the evolution equation 
\Eq{kraichfinal} in \Eq{KLtrans}, and note that 
the eigenfunctions of ${\cal L}$ are given by $\tilde{M}_{\mu}$ 
(in the Landau-Lifschitz approximation),  
with a real eigenvalue $-\gamma$. Then
the function $g$ is given by 
\be
g(\mu,\bar{t}) = g(\mu,0) e^{-\gamma(\mu)\bar{t}}
= e^{-\gamma(\mu)\bar{t}} 
\frac{2\mu \sinh(\pi\mu)}{\pi^2}
\int_0^\infty \frac{dk'}{k'}  K_{i\mu}(k'/k_d){k'}^{3/2}\hat{M}(k',0).
\label{gsoln}
\ee
Here $g(\mu,0)$ is the initial value of $g$, and has itself  
been obtained by evaluating \Eq{invtrans} at $\bar{t}=0$,
and $\hat{M}(k,0)$ is the initial scalar spectrum.
Substituting $g(\mu,\bar{t})$ from \Eq{gsoln} in to \Eq{KLtrans},
allows one to write down the
Green's function solution for the evolution of $\hat{M}(k,t)$
as
\be
\hat{M}(k,t) = k^{-3/2} \int_0^\infty d\mu \  e^{-\gamma(\mu)\bar{t}} \ 
\frac{2\mu\sinh\pi\mu}{\pi^2}  
\int_0^\infty \frac{dk'}{k'} \  K_{i\mu}(k/k_d)K_{i\mu}(k'/k_d) \
{k'}^{3/2} \hat{M}(k',0).
\label{MktGF}
\ee
We note that this solution is identical to the solution got by \citet{BF99} in
the $\tau\to 0$ limit, but now generalised to case of finite $\tau$.

To study this solution in greater detail, we need the explicit form
of $\gamma(\mu)$. To obtain this relation, it is useful to
replace $\gamma_0\tau$ by $\gamma\tau$ to leading order in $\tau$ 
before inverting \Eq{barnu}. We get
\be
\gamma(\mu) = \bar\gamma + \frac{4}{9} \mu^2 \bar\gamma \ 
\frac{(1 - 9\bar\gamma\bar\tau/14)}{(1 + 2\mu^2 \bar\gamma\bar\tau/7)}.
\label{gammamu}
\ee
Note that $\gamma(\mu)$ increases monotonically with $\mu$ from a
value $\gamma = \bar\gamma$, when $\mu=0$
to $\gamma \to \gamma_u = 14/(9\bar\tau)$ as $\mu\to \infty$.
Thus $\alpha$ never becomes negative as required for the
consistency of the solution.
We also have $\gamma_u/\bar\gamma
= 280/81\bar\tau \gg1$ for $\bar\tau \ll1$.
Substituting this $\gamma(\mu)$ into \Eq{MktGF} we then have
\bea
\hat{M}(k,t) &=& k^{-3/2} e^{-\bar\gamma\bar{t}} 
\int_0^\infty d\mu \  \exp\left(-\frac{4}{9} \mu^2 \bar\gamma \bar{t} \ 
\frac{(1 - 9\bar\gamma\bar\tau/14)}{(1 + 2\mu^2 \bar\gamma\bar\tau/7)}\right)
 \ 
\frac{2\mu\sinh\pi\mu}{\pi^2}  
\nonumber \\
&\times&
\int_0^\infty \frac{dk'}{k'} \  K_{i\mu}(k/k_d)K_{i\mu}(k'/k_d) \
{k'}^{3/2}  \hat{M}(k',0) . 
\label{MktGF2}
\eea

We see that there is an overall $k^{-3/2}$ dependence of the spectrum
independent of $\tau$, and also an overall exponential decay of
the spectrum with a decay rate $\bar\gamma$ (as in the scaling solution).
The remaining $k$ and $\bar{t}$ dependence will come from 
evaluating the integral in \Eq{MktGF2} over $\mu$. 

Note that the $\exp(-\gamma(\mu)\bar{t})$ factor in this integral leads to 
a damping of the integral, which increases with $\mu$. For example, with $\bar\tau=0.1$, the extra damping factor
over and above $e^{-\bar\gamma\bar{t}}$, as $\mu\to \infty$ goes as $e^{-15.6\bar{t}}$, while there is no extra damping at $\mu=0$. 
Meanwhile, from a power speries expansion of $K_{i\mu}(k/k_d)$ 
about $k=0$ \citep{dunster90}, the combination $\sqrt{\mu \sinh(\pi\mu)} K_{i\mu}$ 
is bounded as $\mu$ increases. Then the effect of $\gamma(\mu)$ damping results in the integral over $\mu$, being very rapidly dominated
by contributions from the lower range of $\mu$. In fact as time increases, the range of $\mu$ which contributes significantly
to the integral decreases with time, and is limited to $\mu < \mu_0 \approx 1/\sqrt{\bar\gamma\bar{t}}$,
which at late times can become much smaller than unity.
Further we have $K_{i\mu}(k/k_d) \propto \sin(\mu \ln(k/2k_d) + c_\mu)$,
for $k/k_d \ll 1$. Thus, due to the weak logarithimc
dependence on $k$ compounded by the fact that $\mu < \mu_0 \ll1$
at late times, the integral has a weak $k$ dependence below the 
dissipative scales at late times. As a result, the 
$k$ dependence of the scalar spectrum coming from the integral
can become sub dominant to the overall $k^{-3/2}$ dependence
of the spectrum coming from outside the integral. This obtains
for a generic initial spectrum and is also independent of $\bar\tau$. 
On general grounds, we see from \Eq{MktGF2}, 
the scalar spectrum during decay in the Batchelor regime,  
has a dominant form $\tilde{M} \propto k^{-3/2}$
or $E_\theta(k) \propto k^{1/2}$ at late times, independent of $\bar\tau$. We will see these features
more explicitly in the worked example below.

\subsubsection{Numerical integration for scalar spectral evolution}
\label{example}

One can integrate \Eq{MktGF2}
numerically, to determine the scalar spectral evolution
for a fixed $\bar\tau$ and any initial spectrum.
In order to get some physical insight, we focus on
a simple example of the initial spectrum, which nevertheless illustrates
all the features of interest.  We choose a case where we assume that the initial spectrum is very strongly peaked
at some small $k_0/k_d \ll 1$; specifically
that $\hat{M}(k,0) = \delta(k-k_0)/(4\pi k^2)$ 
such that the total initial scalar power 
$\int 4 \pi k^2\hat{M}(k,0) = 1$. We ask how this
evolves in time? Substituting the above form of the initial spectrum
allows one to do the $k'$ integral in \Eq{MktGF2} trivially.
In the small $k/k_d\ll1$ limit, one can also do the $\mu$ integral analytically in an approximate manner,
and we discuss that in Appendix~\ref{solved}. Here we
do not make any approximations and simply do the $\mu$ integration
in \Eq{MktGF2} numerically using Mathematica.

\begin{figure}
\centering
\epsfig{file=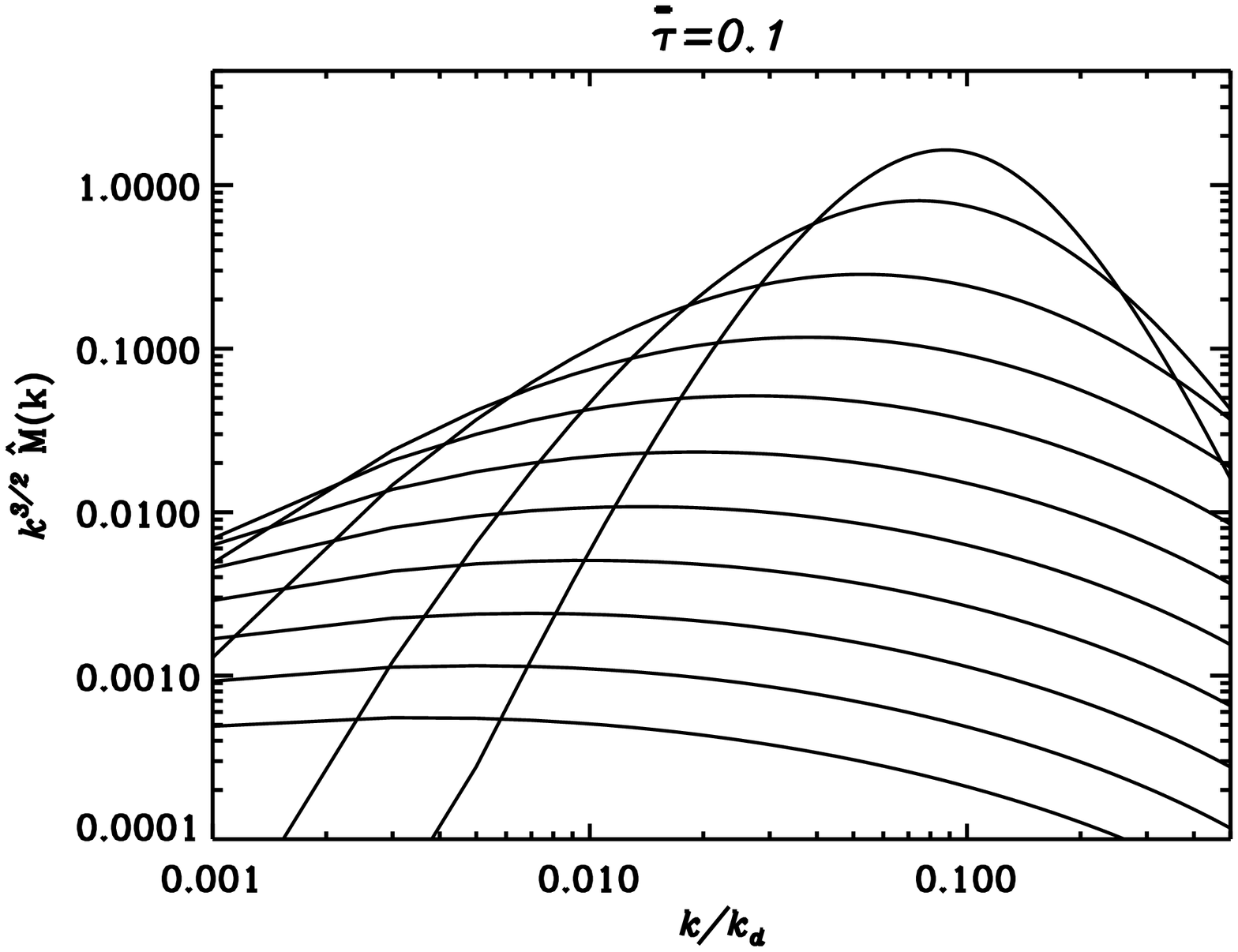, width=0.48\textwidth, height=0.25\textheight}
\epsfig{file=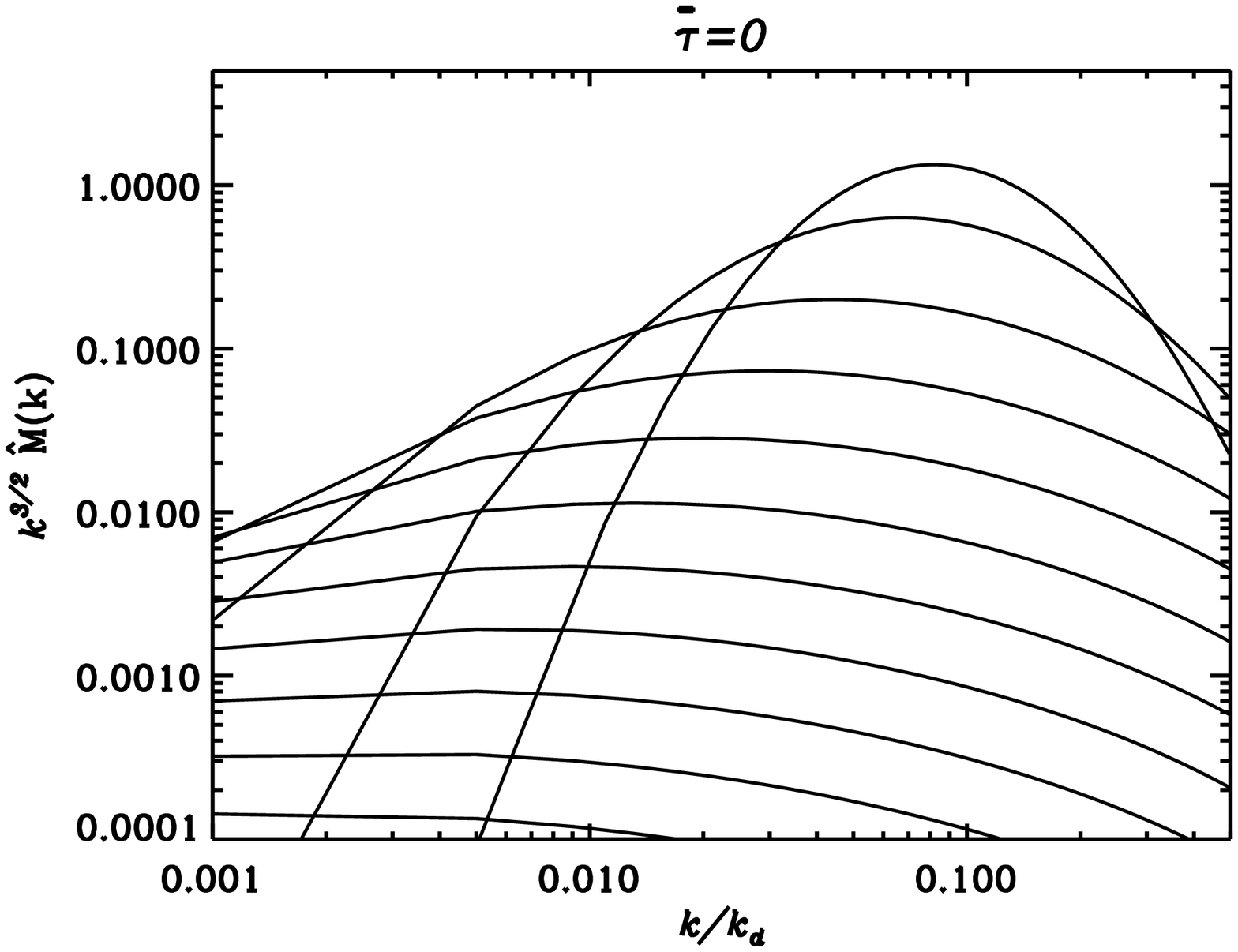, width=0.48\textwidth, height=0.25\textheight}
\caption{
The left and right panels show the decaying scalar spectrum,
multiplied by $k^{3/2}$ factor, for the cases of $\bar{\tau}=0.1$ 
and $\bar{\tau}=0$ respectively. 
The wavenumber $k$ here is measured in units of $k_d$.
Initially power is peaked at $k_0/k_d=0.1$. The topmost curve in each case shows the spectrum at $\bar{t}=1$ or 
one turbulent diffusion time after the decay began.
Subsequent curves, from top to bottom, are shown for times 
$\bar{t}=2, 4, 6, 8....20$. 
A flat curve corresponds to the 3-D spectrum behaving as $\tilde{M}(k) \propto k^{-3/2}$ or
a 1-D spectrum $E_\theta(k) \propto k^{1/2}$.
}
\label{decaymodel}
\end{figure}

In Fig.~\ref{decaymodel} we show the evolution of scalar spectrum multiplied by a $k^{3/2}$ factor, to compensate
the overall $k^{-3/2}$ factor in \Eq{MktGF2}. 
The left and right panels correspond to the cases of $\bar{\tau}=0.1$
and $\bar{\tau}=0$ respectively. 
The decay starts from the initial power peaked at $k_0/k_d=0.1$.
A flat region of the curves in this figure would correspond to where 
the 3-D spectrum behaves as $\tilde{M}(k) \propto k^{-3/2}$ or
a 1-D spectrum $E_\theta(k) \propto k^{1/2}$.
The topmost curve shows the compensated spectrum at $\bar{t}=1$ or 
one turbulent diffusion time after the
the decay began. Subsequent curves, from top to bottom, are shown for times 
$\bar{t}=2, 4, 6, 8,...20$. We see from Fig.~\ref{decaymodel}, that the compensated scalar power
spreads with time, as it decays, 
to wavenumbers both smaller and larger than $k_0$.
The initial delta function 
is already broadened significantly due to turbulent diffusion by $\bar{t}=1$. 
The spectrum cannot spread to large $k > k_d$ due to
diffusion damping. However, it can spread to arbitrarily
small $k/k_d \ll 1$ in the Batchelor regime, as time increases. We see that as this happens, the compensated power spectrum
becomes flat for a range of wavenumbers at the low $k$ end, and this range keeps increasing secularly with time. 
At much smaller $k$ (beyond what has been plotted for) the spectrum is expected to be cut-off as the scalar power has not yet spread to those wavenumbers. 
And at large $k$ there is a faster cut off due to diffusion.
It is for a set of intermediate range of wavenumbers, in the Batchelor regime,
that one would see $E_\theta(k) \propto k^{1/2}$ behaviour of the
scalar spectrum during decay. 
More generally, we note by comparing 
the left and right panels of the Fig.~\ref{decaymodel}, that
the qualitative form of the scalar spectrum is independent of $\bar\tau$.
However, Fig.~\ref{decaymodel} also shows that scalar power spectrum for the $\bar{\tau}=0.1$ case decays slower than $\bar{\tau}=0$, as expected.

Note that although we started with a specific initial spectrum,
the qualitative features of the scalar spectrum during decay 
for $k/k_d\ll1$ is expected to be the same, for any peaked initial
spectrum.

Once the scalar power spectrum has spread to scales
larger than the  forcing scale (to $k < q$), the slower decay of the
larger scale modes by turbulent mixing, could influence the 
decay of scalar power even for $k > q$ (see Section~\ref{decaydns}).
It could lead a more complicated behaviour of the decay phase (see \citet{SHC04} for a model
problem), than what obtains in the purely 
Batchelor regime which is the focus of the present work.
Further theoretical study, also taking into account the large-scale
(small $k$) cut off of the velocity correlations, is important
and will be useful to improve the understanding of the scalar
decay, in both the Kraichnan limit and its finite $\tau$ extensions. We now consider to what extent 
our theoretical predictions are borne out in 
direct numerical simulations of passive scalar mixing
and decay.

\section{Direct numerical simulations of passive scalar evolution}

To perform the direct numerical simulations (DNS) of the passive 
scalar mixing in forced turbulence, we have used 
\textsc{Pencil Code}\footnote{https://github.com/pencil-code} \citep{BD02,B03}.
The fluid is assumed to be isothermal, viscous and mildly compressible.
We solve the continuity, Navier-Stokes and passive scalar equations given by,
\begin{eqnarray}
&& \frac{\DD}{\DD t} \ln{\rho} = -\nab\cdot\uu, \label{eq: continuity} \\
&& \frac{\DD}{\DD t} \uu = -\cs^{2}\nab \ln{\rho} +
\FF_{\rm visc} + \ff,
\label{eq: momentum} \\
&& \frac{\partial}{\partial t} (\rho\theta)= - {\bf \nab} \cdot \left[ \rho\theta\uu-\rho\kappa\nab\theta\right]
\label{eq: pscal}
\end{eqnarray}
Here $\rho$ is the density related to the pressure by $ P=\rho c_s^2$, where $c_s$ is speed of sound. 
The operator $D/Dt=\partial/\partial t + {\bf u}\cdot {\bf \nabla}$ is 
as before the Lagrangian derivative.
The viscous force is given by, 
\begin{equation}
\FF_{visc} = \nu \left[ \nab^2  \uu + \frac{1}{3} \nab\cdot\nabla \uu 
+ 2S\cdot\nab ln \rho\right]
\end{equation}
where,
\begin{equation}
S=\frac{1}{2}\left(\frac{\partial u_i}{\partial x_j}
+\frac{\partial u_j}{\partial x_i}-\frac{2}{3}\delta_{ij}\nab\cdot\uu\right),
\end{equation}
is the traceless rate of strain tensor.
To generate turbulent flow, a random force, $\ff=\ff({\bf x}, t)$, is included manifestly 
in the momentum equation. 
In Fourier space, this driving force is transverse
to the wave vector ${\bf k}$ and localized in wave-number space about
a wave-number $k_f$, driving vortical motions in a wavelength
range around $2\pi/k_f$, which will also be the
energy carrying scales of the turbulent flow. 
The direction of the wave vector and and its phase are changed at every time step in the simulation
making the force almost $\delta$-correlated in time
(see \cite{HBD04} for details).
These equations are solved in a Cartesian box of a size $l=2\pi$ 
on a cubic grid with 
$N^3$ mesh points, adopting periodic boundary conditions.
\begin{table}
\begin{center}
\begin{tabular}{lrlcccc}
\hline
\hline
Run & $\;R_\kappa\!\!$ & $Sc$ & $u_{rms}$ & $k_f$ &$k_0$&$N^3\;\;$ \\
\hline
A &4000  & 400  &  0.15  &  1.5 &-&$1024^3$ \\ 
B &275  &  100  &  0.11  &  4.0 &-& $1024^3$ \\ 

C &3730 & 400 & 0.14 & 1.5 & 30 & $1024^3$ \\ 

\hline
\label{Runtable}
\end{tabular}
\caption{
Summary of runs discussed in this paper.
}
\end{center}
\end{table}

\subsection{Steady state case}

We have used a resolution of $N=1024$ for two simulations choosing 
$k_f=1.5$ (run A) and $k_f=4$ (run B) respectively. 
The basic parameters of the simulations are summarized in Table~\ref{Runtable}.
The initial velocity is zero and the form of initial passive scalar 
is Gaussian random noise, with an arbitrary amplitude of 0.5. 
In order to get a steady state for the passive scalar evolution, 
we impose a constant gradient 
of the passive scalar in an arbitrarily chosen direction 
which is along the diagonal of the box.
This corresponds to a force $f_\theta = \uu\cdot{\bf \Gamma}$ with
a constant ${\bf \Gamma}$, and is a standard technique for 
achieving a steady state; it retains
homogeneity but breaks isotropy at the largest scale.
\begin{figure}
\centering
\epsfig{file=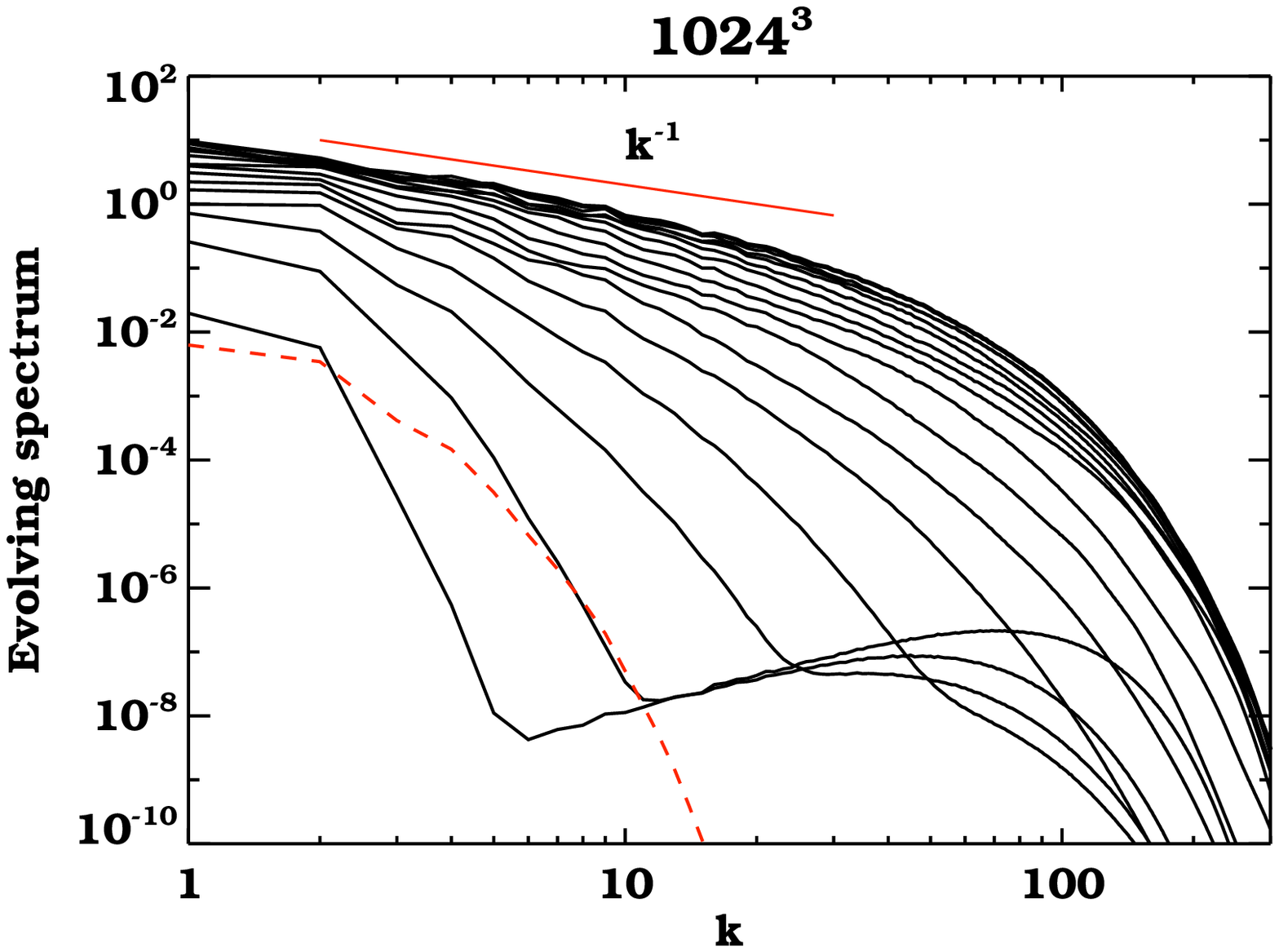, width=0.48\textwidth, height=0.27\textheight}
\epsfig{file=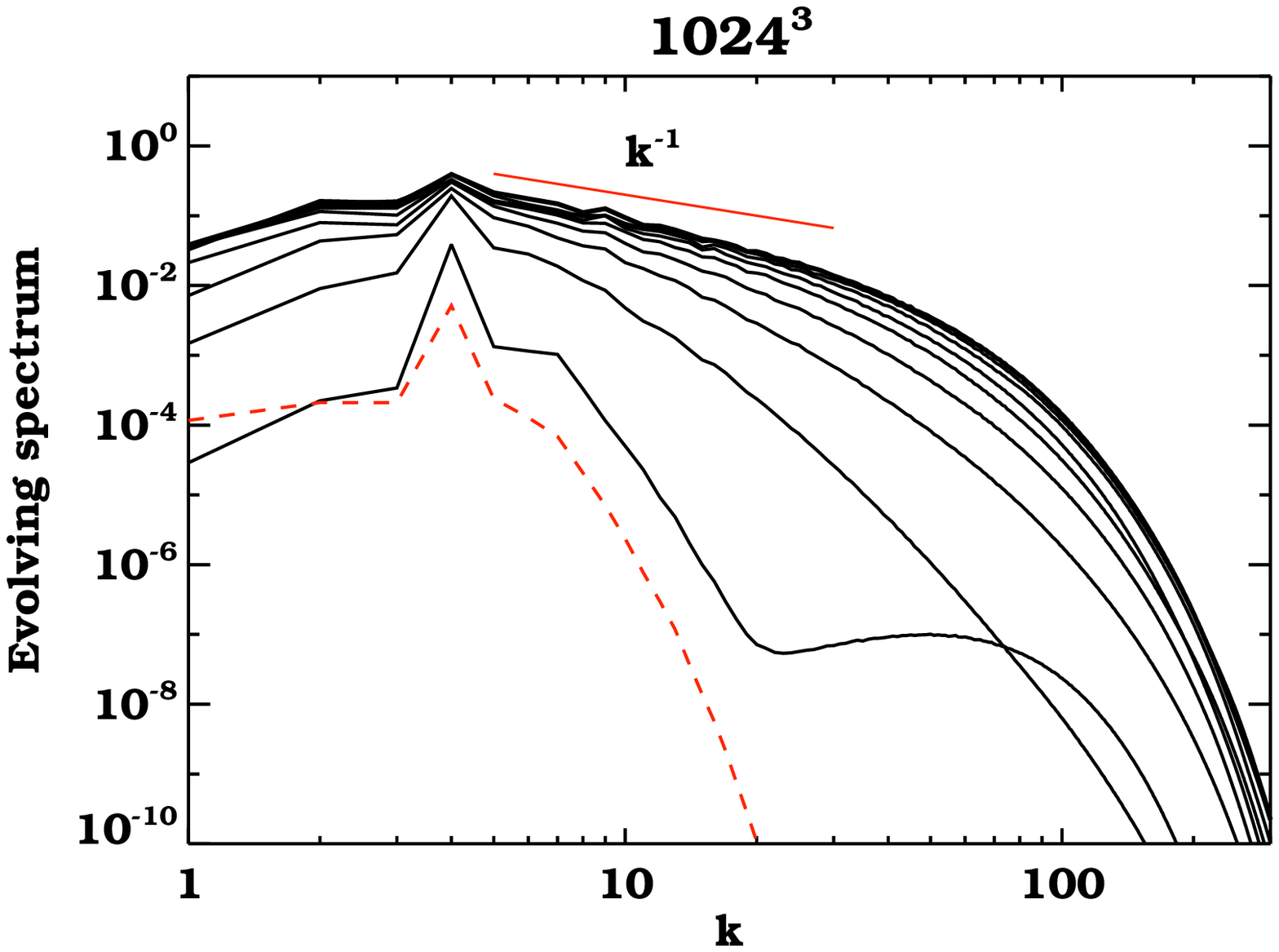, width=0.48\textwidth, height=0.27\textheight}
\caption{The left panel shows the growing scalar spectrum as
solid lines for run A
from times $t=5-80$ every $5$ time units. 
Time increases from the bottom to the top curve and the topmost
curve shows the steady state scalar spectrum.
The steady velocity spectrum
is shown as a dashed line. 
Within $t=20$ the rms velocity rises to $u_{rms} \sim 0.15$ and
varies between $0.13-0.17$, of the sound speed.
The right panel shows the corresponding 
spectra for run B.
The scalar spectra are shown from $t=5-45$ every $5$ time units. 
Within $t=10$ the rms velocity settles to $u_{rms} \sim 0.11$ 
in units of the sound speed.
}
\label{steady1}
\end{figure}

In Fig.~\ref{steady1}, we show as solid lines, 
the evolution of the passive scalar power spectrum, $E_{\theta}$, in equal time intervals. 
The steady state velocity spectrum is also shown, by a dashed line. The left panel is for run A with $k_f=1.5$ and the right panel is 
for run B with $k_f=4$. The rms velocities in units of the sound speed  are $u_{rms}\sim 0.15$ and $u_{rms}\sim 0.11$ in
runs A and B respectively, such that the incompressibility condition is
well satisfied, and $\rho$ is nearly constant. 
The viscosity has been set high enough that 
the flow is almost single scale, and the diffusivity low enough that
the Peclet and Schmidt numbers are high.
We have $Sc=400$, $R_\kappa \sim 4000$ and $Sc=100$, $R_\kappa =250$ for the runs A and B respectively.
Thus one would expect indeed to obtain a significant viscous-convective
range of wavenumbers.  
the simulations are then suited to test if one obtains the
Batchelor $E_\theta \propto 1/k$ spectrum, in steady state.
The simulation were run for a sufficiently 
long time to obtain steady state in evolution of the passive scalar,
as can be seen from the overlap of the spectrum at final times. 
As can be seen from both the panels in 
Fig.~\ref{steady1}, $E_{\theta}$ reaches a steady state
and exhibits a scale dependence of $k^{-1}$ to a reasonable
degree in the viscous-convective range larger than $k_f$. 
While Kraichnan's idealistic model of delta-correlation predicted $k^{-1}$
spectrum, the solution to our extended Kraichnan equation showed 
that to the leading order, finite time correlation 
does not change this scale dependence.
This is also predicted by the Lagrangian analysis \citep{FGV01}.
We find that our DNS which explores Schmidt numbers up to $400$ 
bears this out. We note that earlier DNS,
with higher $\Rey$ than obtained here, but with $Sc$ up to 64, have
also found evidence for the Batchelor spectrum in the viscous-convective
range (see \citet{DSY10,GY13} and references therein). During the
course of this work, we also came across some recent 
very high resolution hybrid simulations of passive scalar mixing 
with $Sc\sim 200-1000$ comparable to ours, which also reports 
reasonable agreement with the steady state Batchelor spectrum \citep{GWM14}.  
There have been interesting predictions for the behaviour of higher
order correlators in the steady state and the influence of the diffusive 
scale even on larger scales \citep{BCKL95,CKL07}. It would be of interest
to check for such effects in future work.
\begin{figure}
\centering
\epsfig{file=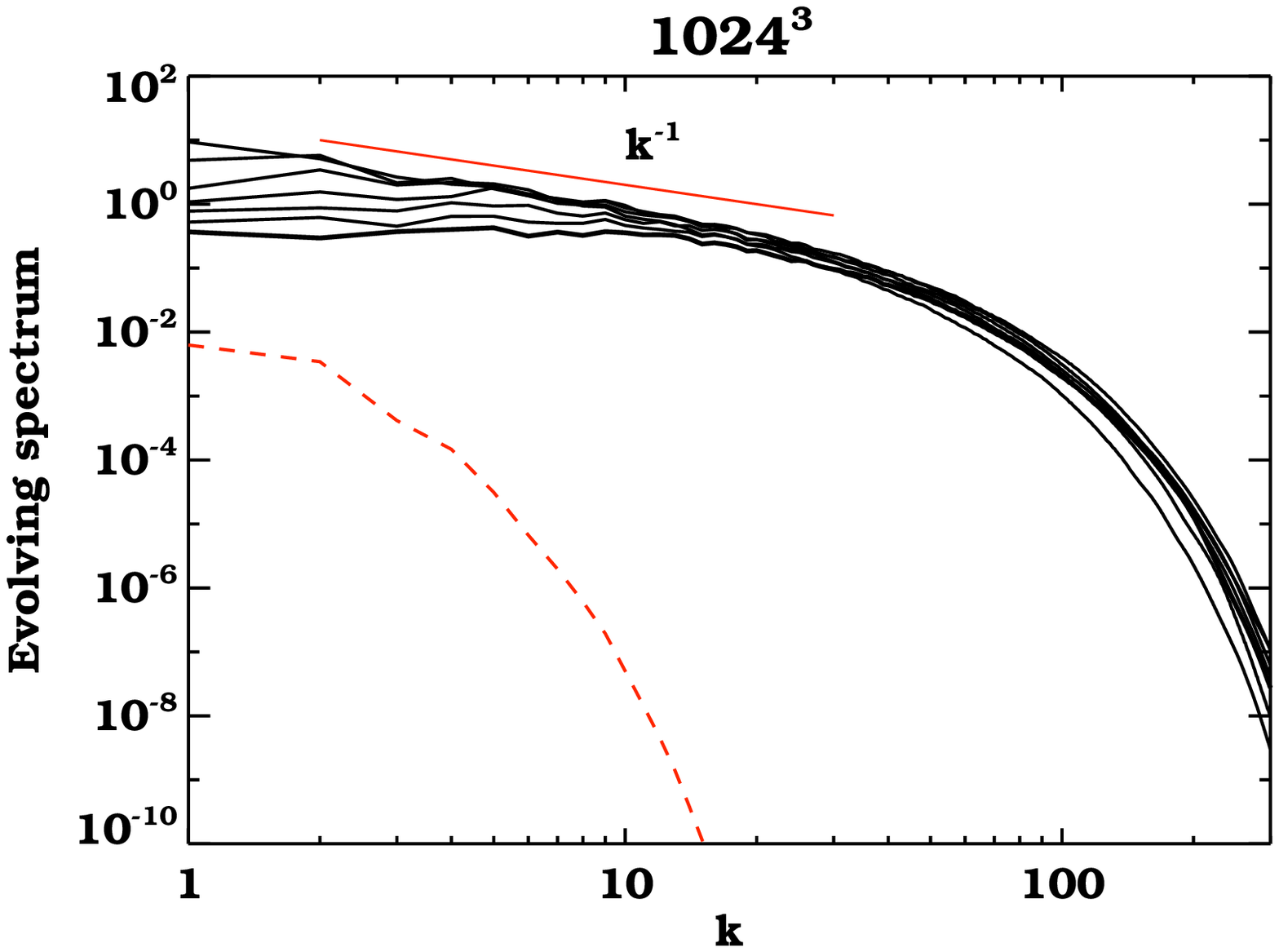, width=0.48\textwidth, height=0.27\textheight}
\epsfig{file=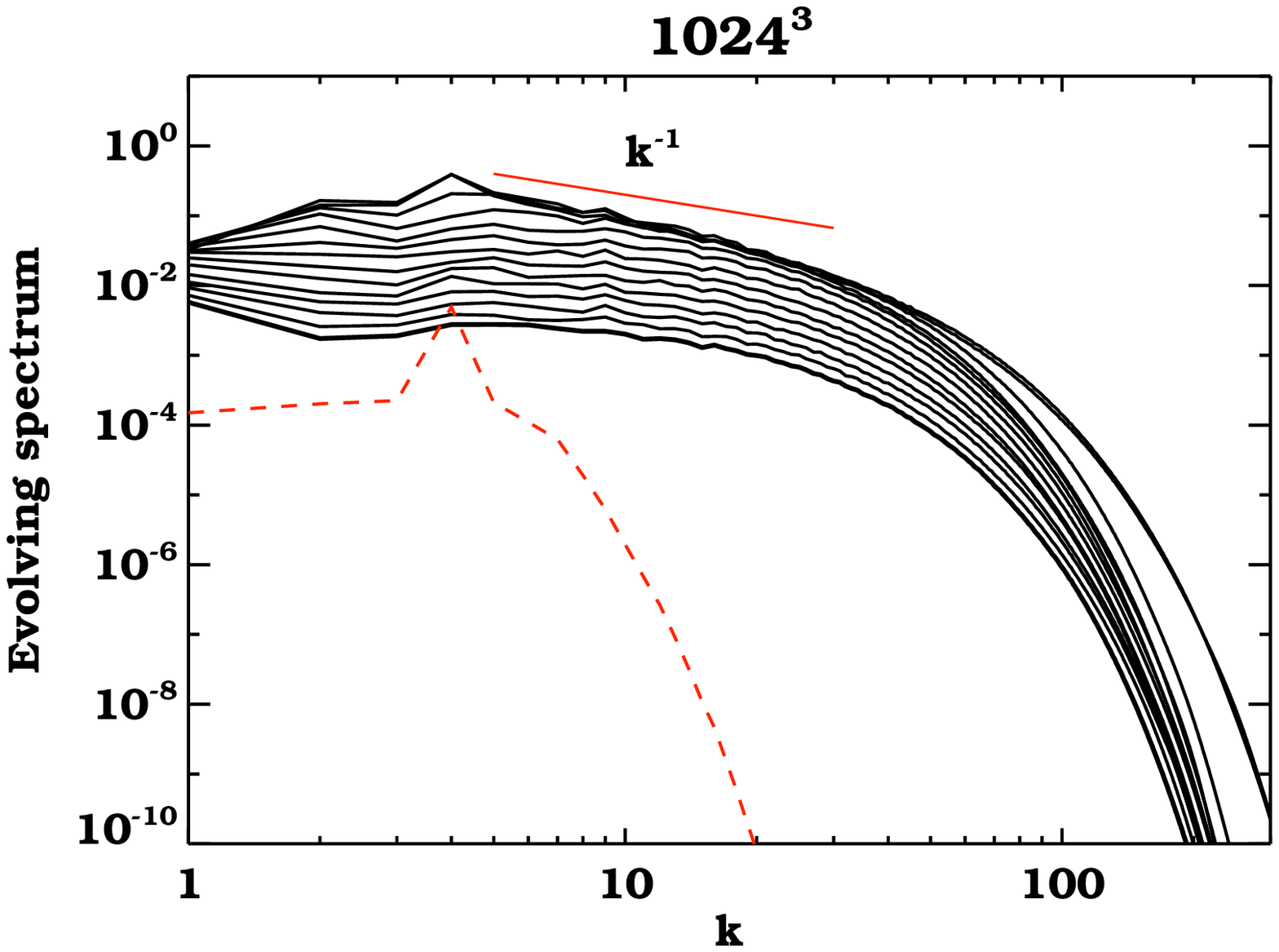, width=0.48\textwidth, height=0.27\textheight}
\caption{The left panel shows the decaying scalar spectrum (solid lines)
from $t=80-140$ in steps of $10$ time units for run A. The corresponding
decay phase for run B is shown on the right panel, 
from $t=40-105$ in steps of $5$ time units. 
Time now increases from the top to the bottom curves.
The topmost curve again shows the steady state spectrum from which
the decay began, once the scalar forcing is turned off.
The dashed lines again show the
steady state velocity spectrum.
}
\label{decay1}
\end{figure}

\begin{figure}
\centering
\epsfig{file=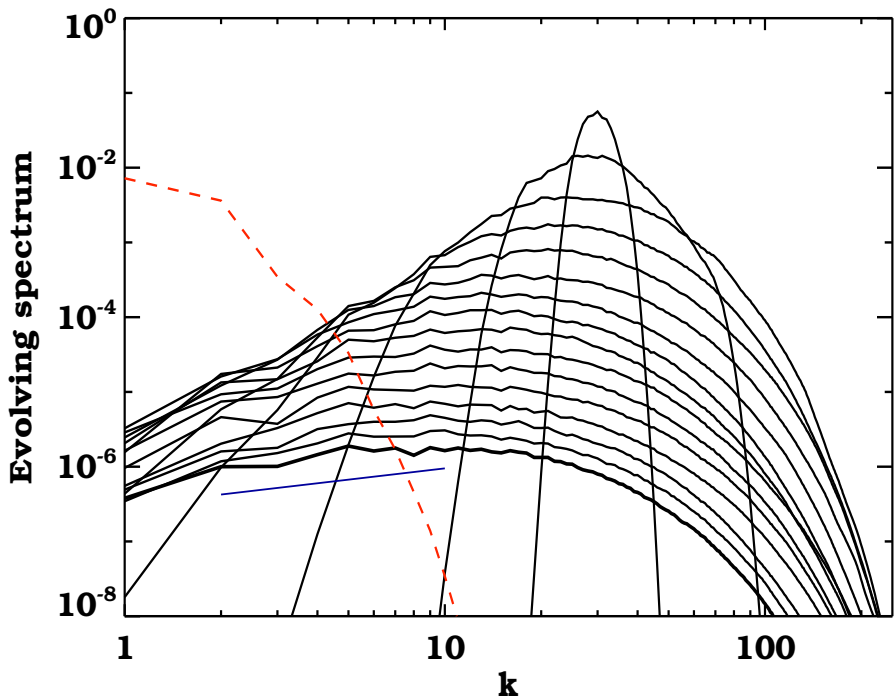, width=0.96\textwidth, 
height=0.48\textheight}

\caption{
The figure shows the decaying scalar spectrum (solid lines)
when one starts from initial power peaked at $k_0=30$.
The topmost curve shows the spectrum at $t=5$ or 
after about 1 eddy turn over time after the
the decay began, with the scalar forcing is turned off.
Subsequent curves, from top to bottom, are shown for times 
$t=5,10,20,.......,150$. 
The blue solid straight line shown 
at the bottom is $E(k) \propto k^{1/2}$.
The dashed line again shows the (almost)
steady state velocity spectrum. 
Within $t=5$ the rms velocity rises to $u_{rms} \sim 0.1$ and
varies between $0.11-0.15$, of the sound speed.
}
\label{decay5}
\end{figure}
\subsection{DNS of passive scalar decay}
\label{decaydns}

We next examine the decay of the passive scalar through our DNS.
We carry out two types of DNS to explore this phase. In the first case, to obtain such a decay, we turn off the imposed constant gradient 
in the passive scalar equation. The passive scalar spectrum then decays from
its initial steady state, as is shown in Fig.~\ref{decay1}. 
We find that the spectrum flattens as it decays, but never becomes
close to the $k^{1/2}$ form predicted by the analytic argument.
This can be seen to obtain in both Run A (left panel of Fig.~\ref{decay1}) 
and in Run B (right panel). It could be that
the box scale mode decays slower due to turbulent diffusion, 
at a rate $\sim k_{box}^2\kappa_t$, with $k_{box} =1$ in our DNS, 
compared to the small scale modes.
Our analytical considerations predict that they decay 
at the rate $\sim q^2 \kappa_t$, with $q\sim k_f$. 
Some evidence for such an idea is seen in Run B, where there
is a clear separation of scales. Then the box scale modes
could act as a source for the small scale fluctuations, leading
to a spectrum flatter than $k^{1/2}$ but not as steep as $k^{-1}$.
A similar idea has been explored in \citet{SHC04}. 

In view of the above, we carry out a second type of DNS where we
make sure that there is no large-scale power in the scalar fluctuations
at the initial time. In particular, we start the decay with scalar
power peaked initially at some wavenumber $k_0 \gg k_f$ and let
it decay due to the random motions driven at the forcing wavenumber $k_f$.
This is analogous to the case of the solved example, presented
in Section ~\ref{example} and appendix~\ref{solved}. 
In Fig.~\ref{decay5}, we show the results from such a 
high resolution run ($N=1024$, Run C), where 
the initial scalar spectrum is a delta function at $k_0=30$ and $k_f=1.5$. 
The subsequent decay of the scalar spectrum is shown as a 
sequence of solid curves in Fig.~\ref{decay5}.
The curves from top to bottom correspond to times $t=5,10,20,30,......,150$. 
The blue straight line is the 1-D scalar spectrum which is
predicted at the low k end in the Batchelor regime,  
$E_\theta(k) \propto k^{1/2}$. 

The decay now is more in line with expectations.
By $t=5$, which is roughly one eddy turn over time, the spectrum
has spread to have almost a Gaussian shape, although
still peaked at around $k\sim k_0=30$. As the decay proceeds, 
the scalar power spreads, with its half width increasing in time. 
The slope at low $k$ is always positive now, but 
steeper than the $k^{1/2}$ form 
because of the influence of the expected rise of power towards $k_0$,
the Gaussian part in \Eq{soln1}.
Power reaches the diffusive cut-off scale $k_d \sim 90$, by $t=20$,
after which it cannot spread further to larger $k$, but continues
to spread to smaller and smaller $k<k_0$. The peak of the spectrum thus shifts
to smaller and smaller $k< k_0$ as the scalar power decays.
The power at $k=1$ increases till about $t=60$, stays roughly constant 
and later after $t=90$ starts decaying. In fact after about $t=90$ 
till $t=140$, the spectra shift down almost like an eigenfunction, 
preserving their shape. 
Towards the end of the simulation, the low-$k$ slope
of the scalar spectrum approaches the $k^{1/2}$ form 
(see Figure~\ref{decay5}), 
with the peak of the spectrum at $k\sim 10$. Eventually, the box scale modes
are expected to decay slower than the modes with $k >k_f$ and 
one could see a further flattening of the spectrum.
All in all this DNS shows that the theory of scalar decay developed
for the Kraichnan model and its generalization,
in Section~\ref{scaldecay}, is borne out to a reasonable extent. 

\section{Discussion and Conclusions}

The mixing and decay of a passive scalar by random or turbulent flows
which are spatially smooth on scales where the scalar still has structure, 
is important in several contexts in nature and industry. 
This regime obtains when the Schmidt number (the ratio of viscosity
to scalar diffusivity) is large; for example
$Sc\sim 10$ for flow of heat in water and of order $1000$ for several
organic liquids \citep{GY13}. The passive scalars then develops
finer scale structure than the velocity itself. \citet{B59} developed a theory for the 
steady state spectrum of fluctuations generated in such mixing, and argued that $E_\theta(k) \propto 1/k$,
a spectral form known as the Batchelor spectrum. This spectrum
was shown to obtain as a steady state solution to the evolution equation
for $E_\theta$ by \citet{K68}.

Kraichnan's theory of passive scalar mixing and decay in such flows assumed delta-correlation in time. 
However, the correlation time is expected to be finite in realistic random flows. We have extended here the study of passive scalar mixing
and decay to finite correlation times, by using a model of random single scale  velocity field that renews itself after every time interval $\tau$.
In particular, we present a detailed derivation of the evolution equation for the power spectrum and the 
two-point correlation of the passive scalar in renewing flows. For this purpose, we used an ensemble average over the random flow,
following one renewing step at a time. We arrived at the generalised Kraichnan equation which includes leading 
next order contributions in $\tau$ when compared to the original Kraichnan 
equation itself. As we explicitly point out above, this equation involves fourth order velocity correlators, which for our flow, are not just
products of the two-point velocity correlator. We note that our differential 
equation approach to finite $\tau$ effects is essentially new, 
and complementary to the 'integral' Lagrangian approaches to the same problem \citep{FGV01}, 
reviewed in Appendix-\ref{appA}.

The generalised Kraichnan equation in Fourier space derived here,
applicable to the viscous-convective range, contains new contributions 
which involve higher order (third and fourth) derivatives 
of the passive scalar power spectrum $\hat{M}(k,t)$. 
However, these higher order terms appear only perturbatively in $\bar\tau = \kappa_t q^2\tau$.
As in BS15, one can then use the Landau-Lifshitz approach, 
earlier used to handle the radiation reaction force, to
rewrite these terms, to involve those with at most 
second derivatives of $\hat{M}(k,t)$.

The resulting evolution equation is analyzed, in terms of eigenmodes,
using both a scaling solution valid in the viscous-convective range, and also a more exact
solution including diffusive effects. Two cases are considered.
First, the steady state situation which one would obtain if there was also
a source of passive scalar fluctuations at low wave numbers, that
maintains the scalar fluctuations against decay.
Second, the turbulent decay case, where there is no such source.

An important consequence of the generalized Kraichnan equation is
that, its steady state solution for the passive scalar spectrum 
in the viscous-convective regime, 
still preserves the Batchelor form $E_\theta(k) \propto 1/k$,
for a finite $\tau$ to leading order in $\tau$.
In fact the generalized Kraichnan equation (\ref{kraichLL})
for the steady state scalar spectrum $\tilde{M}(k)$, 
which takes into account the effect of a finite $\tau$, 
is exactly the same as that derived under the delta-correlation
in time approximation by \citet{K68}.
To some extent, this result is expected from both the work of B59 
and the several subsequent works using a Lagrangian analysis 
of the general finite $\tau$ case (cf. \citet{FGV01} and references therein;
see also Appendix~\ref{appA}). However, our approach in terms
of the generalized Kraichnan equation is different and complementary 
to the earlier work. It also allows for a non-Gaussian velocity field.

A new result of our work, concerns the scalar spectrum for the decaying case,
when there is no source. This is analyzed in terms of eigenmodes
and the corresponding Green's function solution to the initial value problem.We have shown here that 
at late times, the qualitative form of this spectrum is also
independent of $\tau$ to leading order in $\tau$, and
in the Batchelor regime, has a dominant form $E_\theta \propto k^{1/2}$.
This result, is found by \citet{V06} in the Lagrangian approach for the Kraichnan problem with $\tau=0$. However, for a finite $\tau$, it is
not apparent in Lagrangian approaches, as it requires an explicit evaluation of the probability function $P(X)$ at finite $\tau$, which is non trivial.
It illustrates the usefulness of our complementary, differential equation approach to the finite $\tau$ passive scalar problem.
We also showed that the scalar fluctuations seeded at some initial $k_0$ spread
in time and in addition, have an overall exponential decay at the turbulent 
diffusion rate. An exponential decay of the scalar variance in the Batchelor limit of the
Kraichnan equation has been pointed out earlier \citep{BF99,Son99}. 
The decay rates are however reduced for a finite $\tau$ compared to the delta-correlated flow.
More work incorporating a large-scale cut-off in velocity correlations would be important to understand scalar decay, not only in the
Kraichnan limit, but also its extensions to the finite $\tau$ case.

Further, to test the analytics we have 
carried out high resolution ($1024^3$) DNS of passive scalar mixing at high Schmidt number up to 400,
and Peclet numbers 10 times larger. Such high Schmidt and Peclet numbers are 
comparable to the highest values realized so far, for examining
the validity of the Batchelor steady state spectrum. Our DNS results lend reasonable support to the Batchelor $E_\theta \propto 1/k$ 
scaling when a steady state is maintained by a low wavenumber source.

Moreover, we have also used high resolution DNS to explore  
the turbulent decay case, when there are no sources of scalar fluctuations.
We have examined decay from two types of initial conditions.
In the first case, we start the decay from the steady state spectrum
by turning of the scalar source. In this case, the passive scalar spectrum during 
decay, flattens from the Batchelor $E_\theta(k) \propto (1/k)$ form.
However it does not become as steep as the  
$k^{1/2}$ spectrum predicted by both the Kraichnan model
and its finite $\tau$ generalization presented here. 
This result which is new could arise due to the fact that lower wavenumber, box scale, scalar fluctuations decay at a smaller 
rate, $k_{box}^2\kappa_t$, compared to the small scale ones, which
decay at the rate $\sim k_f^2\kappa_t$. They can then continue to source
the small scale fluctuations, and lead to a shallower spectrum.

We have also carried out DNS of decay where the scalar spectrum
is initially peaked at some wavenumber $k_0\gg k_f$, such that the
box scale modes $k<k_f$ are grossly sub dominant.
The decay behaviour from the DNS is now more in line with expectations
from the analytics. The scalar power spreads as it decays, with the spectrum
having a positive slope at low $k$ at all times. 
At late times, when the range over which
the scalar spectrum extent becomes large enough, one sees the emergence of 
$E_\theta \propto k^{1/2}$ behaviour in the Batchelor regime, 
at $k> k_f$. At even later times, one would expect the box scale power (which decays slower) to
become important, and a further flattening of the spectrum as described above.

We used the renewing flow to derive the generalized Kraichnan equation;
however the resulting evolution equation can be written in terms of
velocity correlators $T_{ij}$ and $T_{ijkl}$ in a general form.
It also matches exactly with the corresponding real and Fourier space
Kraichnan equations derived assuming delta-correlated flow in the
appropriate limit of $\tau\to0$. Thus they could have more general validity
than the context of renewing flows, in which they are derived. 
An analysis of this finite $\tau$ equation 
incorporating both scale dependent
velocity correlators and correlation times 
(cf. \citet{CFL96,EX2000,AAH02,CGHKV03}) would be useful.
It would also be of interest to try and extend our results to
the non-perturbative regime, through both the current approach and a Lagrangian analysis.
In addition, it will be fruitful to 
include the effects of shear and
rotation, which are often present in realistic turbulent flows.

\acknowledgments
AKA thanks Prof. Rama Govindarajan at TCIS, Hyderabad
for support during the final stages of this work. 
PB thanks Dr. Fatima Ebrahimi for support under DOE grant 
DE-FG02-12ER55142. We thank Dr. Ryan White for timely 
help with integration tools.The simulations presented here used the IUCAA HPC.
We thank the referees for comments which helped 
to improve the paper. One referee in particular is thanked
for leading us to improve our discussion of scalar decay.
\appendix
\section{Lagrangian analysis of Passive scalar mixing}
\label{appA}

We review here a Lagrangian analysis for passive scalar mixing
following to some extent 
\citet{FGV01}. This will also clarify the conditions under which one can derive from such an analysis,  the Batchelor spectrum or the scalar
spectrum when its fluctuations decay, for a general renovation time $\tau$.

Adding a source term to \Eq{scalareqn} for the evolution of the passive scalar field $\theta(\x,t)$ gives, 
\begin{equation}
\frac{\partial \theta}{\partial t} + \uu\cdot{\bf \nabla}\theta =
\frac{D\theta}{Dt} = f_\theta(\x, t) + \kappa\nabla^2 \theta,
\label{sourceeqn}
\end{equation}
Suppose we consider scales where diffusion can be neglected.
Then the formal solution to \Eq{sourceeqn} is
\be
\theta(\x,t) = \theta_0(\xo,t_0) 
+ \int_{t_0}^t f_\theta(\x(t'),t') dt'
\label{formal}
\ee
where $\theta_0$ is the initial scalar field and the integration
of the source is along the Lagrangian trajectory of a particle
frozen to the flow. Further $\xo=\x(t_0)$ in $\theta_0$ 
is related to $\x(t)$ in $\theta$ by
integrating back in time along the Lagrangian trajectory.
\subsection{The steady state Batchelor spectrum}
To begin with, let us assume that the initial scalar field fluctuations
are either small or have decayed, and focus on those sustained by the
source term. The two-point spatial correlation function of the passive scalar,
as defined by \Eq{scalarcor}, then evolves as,
\be
M(r,t) = 
\bbra{\int_{t_0}^t dt'\int_{t_0}^t ds
  f_\theta(\x(t'),t') f_\theta(\y(s),s)},
\label{correvol}
\ee
where the source is assumed to be uncorrelated with the
initial scalar field $\theta_0$ and $r=\vert\x-\y\vert$ as before.
The averages over $f_\theta$
involves both average over the source statistics and that over
the random Lagrangian trajectories $\x(t')$ and $\y(t')$.
It is generally assumed \citep{FGV01}, that the source is statistically
homogeneous, isotropic, Gaussian, of zero mean and 
variance which is delta-correlated in time, i.e
\be
\bra{f_\theta(\x,t') f_\theta(\y,s)}
= \Psi\left(\frac{r}{L}\right)\delta(t'-s).
\label{fcor}
\ee 
Here $L$ is some fiducial length scale over which $\Psi$ varies.
Using \Eq{fcor} in \Eq{correvol}, we have
\be
M(r,t) = \bra{ \int_{t_0}^t dt' \ \Psi\left(\frac{r(t')}{L}\right)}
= \int \int_{t_0}^t dt' \ \Psi\left(\frac{r'}{L}\right)  P(r',r;t,t') dr'.
\label{correvol2}
\ee
The averaging in the first part of the equation is
over the random Lagrangian trajectories of a pair of particles
frozen to the flow, whose separation is $r(t')$ at time $t'$. 
In the latter part of the equation this has been 
written in terms of the two-particle pair probability $P(r',r;t',t) dr'$,
that for a given pair separation, $r=r(t)$ at time $t$, the
pair separation is $r'=r(t')$ at an earlier time $t'$.
For finding this probability we have to follow a
pair of particles, backward in time from $t$ to $t'$. 
However, some general results can be obtained even without
finding the exact form of $P$. 

Suppose the flow has a correlation time $\tau$ much smaller than the time 
$T=t-t_0$. Then the Lagrangian trajectory will be comprised of
many uncorrelated steps. For our renewing flow in 
particular, the frozen in pair of 
particle will take $N=T/\tau$ random steps. For a smooth flow,
like the one we consider, in the Batchelor regime, with 
$1/k_\kappa \ll r \ll 1/q$, the mean pair separation is expected to increase exponentially fast, with $r_0/r = \exp(\bar\lambda T)$.
Here $\bar\lambda > 0$ (see below) and 
$-\bar\lambda$ is the Lyapunov exponent for the contracting
direction, when going forward in time 
(cf. Eq. (138) of \citet{FGV01}), and is
negative for an incompressible flow. We will calculate 
$\bar\lambda$ explicitly below in the Kraichnan limit. For sufficiently large $N\gg1$, the peak of the probability 
distribution $P$ then shifts to
$r' = r \exp(\bar\lambda T')$ as $T'= t-t'$ increases backward in time.
We can then approximate the integral in \Eq{correvol2} as follows.

We note that significant contribution to the integral over $r'$
will obtain only for $r' < L$, corresponding to a time $T' < T_* = 
\bar\lambda^{-1} \ln(L/r)$. Let us approximate the value
of $\Psi(r)$ in this range by its value at the origin, $\Psi_0$.
Then
\be
M(r,t) \approx \Psi_0 \int_0^{T_*} dT' =\Psi_0 T_* = \frac{\Psi_0}{\bar\lambda}
 \ln\left(\frac{L}{r}\right).
\label{corrbat}
\ee
The logarithmic form of the two-point spatial correlator in \Eq{corrbat} 
is equivalent to the $E_\theta(k) \propto 1/k$ form of the Batchelor
spectrum. Thus the Batchelor spectrum arises in a fairly general 
manner independent of the value of the renewal time $\tau$.
We emphasize that the above derivation involves some simplifying assumptions:
(1) The source is delta-correlated in time; one could replace this
with a sufficiently short correlation time compared to $T$, such that $N\gg1$.
(2) The spatial variation of $\Psi(r)$ can be neglected in the
Batchelor regime. This is equivalent to assuming that in spectral
space the source contributes only at low $k$ outside the
viscous-convective range of $k$.
(3) The width of the probability distribution $P$ is
narrow enough that only the location of its peak is important
in determining $T_*$. 

In this context, we note that the derivation of the
Batchelor spectrum given in the main text, 
using the extended Kraichnan differential equation \Eq{kraichfinal}, 
does involve making assumption (2) above, but not (1) and (3) in an 
explicit manner. Thus it offers a complementary approach to the problem, albeit
an approach which can at present only recover the Batchelor
spectrum for a finite $\tau$ in the perturbative limit.
  
\subsection{The Lyapunov exponent for the renewing flow}
Let us now turn to an explicit derivation of $\bar\lambda$ 
for renewing flows, in the case of a short renewal time $\tau$ and in the limit of a
linearly varying velocity field.
We expand \Eq{uturbdef} about $\x=0$ keeping
only up to terms linear in $\x$. Thus the velocity field is taken to be
\be
\uu = \aaa (\q\cdot\x) \cos\psi + \aaa\sin\psi
\label{linearu}
\ee
in each renewing flow step, with $\aaa$, $\q$ and $\psi$ having
the same statistical properties as before.

Suppose the time interval $T=t-t_0 = N\tau$, corresponding to $N$ intervals of the
renewing flow. Consider a pair of points frozen to the flow which at
time $t$ have a separation $r=r(t)$ and at time $t_0$ a separation
$r_0 = r(t_0)$. Here $r(t) = \vert \rr \vert$ where as
before $\rr(t) = (\x(t)-\y(t))$.  
Let the pair separation at an intermediate time $t_n= t_0+ n\tau$
be $r_n = r(t_n)$. Thus $n=0$ corresponds to the time $t_0$ and
$n=N$ to the time $t$.
Then
\be
\frac{r_0}{r} = \frac{r_0}{r_1}\frac{r_1}{r_2} .....\frac{r_{N-1}}{r}
\ee
Thus the ratio of the initial to final pair separation is
a product of $N$ random variables. To work out its statistical
properties, it is useful to take its logarithm and consider
this as the sum of N random variables, 
\be
X = \ln\left(\frac{r_0}{r}\right)
= \sum_{n=1}^N \ln \left(\frac{r_{n-1}}{r_n}\right)
= \sum_{n=1}^N x_n
\label{Xdef}
\ee
where $x_n = (r_{n-1}/r_n)$. 
Note that the mean of each $x_n$, defined as $\bra{x_n} = \mu$, 
is identical as each renewing flow step has the same statistical properties.
Then the mean of the random variable $X$ is
$\bra{X} = N\mu$. This can be used to define the Lyapunov exponent 
\be
\bar\lambda = \lim_{t\to \infty} \frac{1}{t} 
\bbra{\ln\left(\frac{r_0}{r}\right)} =
\lim_{N \to \infty} \frac{\bra{X}}{N\tau}
= \frac{\mu}{\tau}.
\label{Lyp}
\ee
(For our renewing flow, as we show below, one gets the same 
Lyapunov exponent $\bar\lambda$ if we calculate the exponential 
increase in pair separation going forward in time).
To calculate $\mu$, we need to relate the pair separation
at time $t_n$ to that at time $t_{n-1}$.  
Now integrating $d\x/dt = \uu$ backward in time for a time $\tau$, 
noting again that $\q\cdot\x$ is constant for an incompressible flow,
the evolution of pair separation is given by, 
\be
\rr_{n-1} = \rr(t_{n-1}) = \rr_n - \tau\aaa(\q\cdot\rr_n)\cos\psi 
\label{trajr}
\ee
for each step $\tau$ back in time. 
Taking the amplitude of the vectors on both sides of \Eq{trajr}, gives
\be
x_n = \frac{1}{2}\ln\left[1 - 2\tau\cos\psi(\aaa\cdot\hat{\rr}_n)
(\q\cdot\hat{\rr}_n)
+\tau^2\cos^2\psi (\q\cdot\hat{\rr}_n)^2 \right]
=\frac{1}{2} \ln [1 + \tau\alpha + \tau^2\beta],
\label{rnratio}
\ee
where we have defined $\alpha =- 2\cos\psi(\aaa\cdot\hat{\rr}_n)
(\q\cdot\hat{\rr}_n)$, $\beta = \cos^2\psi (\q\cdot\hat{\rr}_n)^2$
and $\hat{\rr}_n$ is the unit vector $\rr_n/r_n$.

Note that $\mu=\bra{x_n}$ is difficult to calculate analytically
for a general $\tau$ because all the random variables
over which averages have to be taken are inside a logarithm.
This is the reason it is difficult to calculate $P$ exactly
for a general $\tau$ and also hence the properties of the
decaying scalar spectrum exactly (see below). Recall that estimating
the form of the steady state spectrum
did not require explicit calculation of $\bar\lambda$ or $\mu$.
However, $\mu$ and various
other statistical moments of $x_n$ can be calculated
perturbatively for small $\tau$ by expanding the logarithm in
\Eq{rnratio} as a power series in $\tau$ about $\tau=0$.

It is also interesting to point out that powers of $\tau$
are always accompanied by the same powers of $\cos\psi$. As odd
powers of $\cos\psi$ average to zero, all moments of $x_n$
depend only on even powers of $\tau$. Therefore the Lyapunov
exponent defined by calculating the increase of pair
separation going forward in time, is same as that defined going
back in time. 
Together with the incompressibility condition, 
this also implies that the three Lyapunov exponents
for the renewing flow are given by $(\bar\lambda, 0, -\bar\lambda)$.
 
We calculate here just the lowest order contribution
to $\mu$ which can be obtained by expanding the logarithm in \Eq{rnratio}
to up to $\tau^2$ terms. This gives,
\be
\mu = \frac{1}{2} \tau^2 \bbra{\left(\beta - \frac{\alpha^2}{2}\right)}
= \frac{A^2\tau^2q^2}{30} = \frac{3}{5} \kappa_t q^2 \tau
\label{mu0}
\ee
Thus the Lyapunov exponent $\bar\lambda = \mu/\tau = (3/5) \kappa_t q^2$
is of the order the turbulent diffusion rate.

\subsection{The passive scalar correlations during decay}

We briefly comment on the case when the source is absent.
We saw in the main text that the passive scalar fluctuations decay in this
case, with a characteristic spectrum $E_\theta \propto k^{1/2}$ in the
Kraichnan limit. We also showed that interestingly, the spectral
form $E_\theta \propto k^{1/2}$ is preserved even at finite $\tau$
to leading order in $\tau$, although the decay rate decreases for
a finite $\tau$. One may wonder if such results can and have been
obtained in a Lagrangian analysis.

In the absence of a source, the 
two-point spatial correlation evolves as,
\be
M(r,t) = \bbra{\theta(\xo,t_0)\theta(\yo,t_0)} = 
\int dr_0 \ P(r_0,r;t_0, t) \ M(r_0,t_0).
\label{cordec}
\ee
Here we have assumed that the passive scalar at any time
is statistically isotropic and homogeneous and again 
averaged over the random pairs of Lagrangian trajectories which
reach a separation $r$ at time $t$ starting from a separation $r_0$
at time $t_0$. This is explicitly
incorporated by weighting $M(r_0,t_0)$ by 
the pair probability density $P(r_0,r;t_0,t)$ and integrating over $r_0$.

Note that \Eq{cordec} gives an integral equation for
the spatial correlation $M(r,t)$. Its solution depends
on knowing the explicit form of the probability $P$.
This is unlike the steady state case, when one only needed
to know that particle trajectories separate exponentially
fast in a smooth random flow.
In order to calculate $P$ for a general $\tau$, one would
need to know for example its moments and this in turn
involves all the moments of $x_n$. As we noted above, calculating
the moments of $x_n$ involves carrying out various statistical 
averages over $\aaa$, $\q$ and $\psi$, which are inside a logarithm.
This cannot be done analytically for a general $\tau$, and hence the
difficulty with going beyond the Kraichnan limit for the
decaying case correlators. Indeed due to the logarithmic form
of $x_n$, one expects its statistics and hence $P(x)$ to be strongly
non-Gaussian as well. 

The probability $P(x)$ however becomes Gaussian in $\ln(r_0/r)$ (or log-normal
in $r_0$) in the Kraichnan limit (cf. \citet{FGV01} and references therein).
The mean value of $P(X)$ is given by $N\mu$
and dispersion by $N\sigma^2$, where $\sigma^2$ is the variance
of $x_n$ calculated also perturbatively. In this case, the integral
equation can be analysed to show that $M(r,t) \propto r^{-3/2}$, which
corresponds to the $E_\theta \propto k^{1/2}$ spectrum
(cf. \citet{V06}).
However, the result shown in the main text, that the scalar spectrum 
in this case, is also left invariant for a finite $\tau$ to leading order, does not appear to have
been shown using a Lagrangian analysis. This again illustrates the
usefulness of the complementary, differential equation approach
to the passive scalar problem, in terms of the extended
Kraichnan equation, discussed in the main text. 
We plan to return to a more detailed discussion 
of the Lagrangian analysis for the decaying case in future work,
also including the effects of scalar diffusion neglected above. 

\section{A solved example of scalar spectral evolution}
\label{solved}

To get some further insight into the evolution of the scalar spectrum implied by \Eq{MktGF2}, we consider an analytically solvable example. 
As in the main text, assume $\hat{M}(k,0) = \delta(k-k_0)/(4\pi k^2)$.  We are particularly interested also in the small
$k/k_d\ll1$ behaviour of the spectrum, (as we expect the spectrum to
decay at large $k/k_d\gg1$; see below). 
In this limit, using \Eq{decaysmallx}, and carrying out trivially
the $k'$ integral of \Eq{MktGF2}, the scalar spectral evolution 
is given by
\bea
\hat{M}(k,t) &=& \frac{2}{\pi} k^{-3/2} k_0^{-3/2} e^{-\bar\gamma\bar{t}} 
\int_0^\infty d\mu \  \exp\left(-\frac{4}{9} \mu^2 \bar\gamma \bar{t} \ 
\frac{(1 - 9\bar\gamma\bar\tau/14)}{(1 + 2\mu^2 \bar\gamma\bar\tau/7)}\right) 
\nonumber \\
&\times&
\sin\left(\mu \ln(k/2k_d) + c_{\mu} \right)
\sin\left(\mu \ln(k_0/2k_d) + c_{\mu} \right).
\label{Mktsmallk}
\eea
As discussed above, due to the damping factor $\exp(-\gamma(\mu))\bar{t})$, 
which monotonically increases with $\mu$, 
the $\mu$ integral will be dominated by the
contribution from the vicinity of small $\mu$.
As time increases this range of $\mu$ decreases and will be
confined to $\mu < \mu_0 \approx 1/\sqrt{\bar\gamma \bar{t}}$.

We can then approximate 
$2\mu^2 \bar\gamma\bar\tau/7 \sim \bar\tau/\bar{t} \ll1$,
and $(1 + 2\mu^2 \bar\gamma\bar\tau/7) \approx 1$.
Moreover in \Eq{Mktsmallk}, the product
$2\sin(\mu \ln(k/2k_d) + c_{\mu})
\sin(\mu \ln(k_0/2k_d) + c_{\mu})
= \cos[\mu \ln(k/k_0)] - \cos[\mu\ln(kk_0/4k_d^2) + 2c_\mu]$.
Then \Eq{Mktsmallk} becomes,
\be
\hat{M}(k,\bar{t}) = \frac{1}{\pi} k^{-3/2} k_0^{-3/2} e^{-\bar\gamma\bar{t}} 
\int_0^\infty d\mu \  e^{-\beta\mu^2 \bar{t}}
\left[
\cos\left(\mu \ln\left(\frac{k}{k_0}\right)\right) - 
\cos\left(\mu\ln\left(\frac{kk_0}{4k_d^2}\right) + 2c_\mu\right)
\right],
\label{Mktsmallk2}
\ee
where we have defined $\beta = 4\bar\gamma (1 - 9\bar\gamma\bar\tau/14)/9$.
The integral over $\mu$ involving the first cosine term in
\Eq{Mktsmallk2} can be done exactly, while the presence
of $c_\mu$ prevents one from an exact evaluation of the second cosine term.
However, as $(kk_0/k_d^2)/(k/k_0) = k_0^2/k_d^2 \ll 1$
and also $(kk_0/k_d^2) \ll1$, we will have for any $k$,
$\vert\ln(kk_0/4k_d^2)\vert \gg \vert\ln(k/k_0)\vert$.
The phase of the cosine in the second integral will then vary much
more rapidly compared than the phase of the cosine 
in the first integral, as $\mu$ varies. Thus the second integral
will suffer from much larger cancellations and 
hence is sub-dominant compared to the first. 
Then evaluating the first integral in \Eq{Mktsmallk2}, we get
for the one-dimensional scalar spectrum 
\be
E_\theta(k,\bar{t}) = 
4\pi k^2\tilde{M}(k,\bar{t})=
e^{-\bar\gamma\bar{t}} \left(\frac{k}{k_0}\right)^{1/2}
\frac{2}{k_0} \sqrt{\frac{\pi}{\beta\bar{t}}}
\exp\left(-\frac{[\ln(k/k_0)]^2}{4\beta\bar{t}}\right)
\label{soln1}
\ee
(We note in passing that the second integral can also be evaluated
exactly at late times, as then the integral is dominated 
by $\mu < 1/\sqrt{\bar\gamma \bar{t}} \ll 1$, and 
we can take $c_\mu \approx \bar{c} \mu$. Then the second integral
gives a contribution with $\ln(k/k_0)$ in \Eq{soln1} replaced
by $\ln(kk_0/4k_d^2)+2\bar{c}$. As discussed above, 
$(\ln(kk_0/4k_d^2))^2 \gg (\ln(k/k_0))^2$,
and thus the second term can be seen to be explicitly sub dominant 
compared to the first integral and hence its neglect above justified.)

We see from \Eq{soln1} that the evolution of the scalar spectrum 
seeded at $k_0$ is as expected earlier on qualitative grounds. 
It evolves in time with $E_\theta$ taking the form of a Gaussian in 
$\ln(k/k_0)$, with the Gaussian width increasing as $\sqrt{\beta\bar{t}}$,
and multiplied by an overall factor $\propto (k/k_0)^{1/2}$.
When the width becomes large enough,
we expect the $k^{1/2}$ form of the spectrum to dominate near the
maximum of the Gaussian.  Apart from the spreading of the scalar power, there is also
overall exponential decay at the rate $\bar\gamma$ (or of order the
turbulent diffusion rate in dimensional units). 

Note also that $\bar\tau$ appears only in
$\bar\gamma$ and $\beta$, thus the qualitative form
of the spectrum is independent of $\tau$. This is an
important result that a finite $\tau$ does not qualitatively change the
shape of the scalar spectrum during decay. Further, 
both $\bar\gamma$ and $\beta$ are smaller for $\tau\ne0$
compared to the Kraichnan case of $\tau\to 0$. Thus the 
decay of scalar fluctuations is slowed down for a 
non zero $\tau$. The above analysis is valid provided $k/k_d \ll 1$. Once the
spectrum has spread to diffusive scales with $k/k_d > 1$, 
or for more general initial conditions, 
an analytic evaluation of the integrals in \Eq{MktGF2}
is no longer possible, and we have to treat the problem
numerically, as done in the main text.
\bibliographystyle{jfm}
\bibliography{reftau}
\end{document}